# Trace formulae for three-dimensional hyperbolic lattices and application to a strongly chaotic tetrahedral billiard

by

R. Aurich[1] and J. Marklof[2]

II. Institut für Theoretische Physik , Universität Hamburg
Luruper Chaussee 149 , 22761 Hamburg
Federal Republic of Germany

**Abstract**

This paper is devoted to the quantum chaology of three-dimensional systems. A trace formula is derived for compact polyhedral billiards which tessellate the three-dimensional hyperbolic space of constant negative curvature. The exact trace formula is compared with Gutzwiller's semiclassical periodic-orbit theory in three dimensions, and applied to a tetrahedral billiard being strongly chaotic. Geometric properties as well as the conjugacy classes of the defining group are discussed. The length spectrum and the quantal level spectrum are numerically computed allowing the evaluation of the trace formula as is demonstrated in the case of the spectral staircase $\mathcal{N}(E)$, which in turn is successfully applied in a quantization condition.

[1] Supported by Deutsche Forschungsgemeinschaft under Contract No. DFG–Ste 241/6–1
[2] Supported by Deutsche Forschungsgemeinschaft under Contract No. DFG–Ste 241/7–1

# I Introduction

In the last years great efforts have been undertaken to unravel the mystery of quantum chaos, i. e., the search for the fingerprints of classical chaos left on the corresponding quantum systems. The main focus has been on two-dimensional billiard systems, which constitute the simplest class of systems being strongly chaotic. The chaotic behaviour is classically imprinted in the structure of phase space where topological differences occur between Hamiltonian systems with two degrees of freedom and more than two degrees, e. g., the Arnold diffusion in KAM theory is only possible in three or more degrees of freedom. Thus it is about time to investigate also the quantum mechanics of chaotic systems with more than two degrees of freedom, since the different phase space structures semiclassically influence quantal energy levels as well as the eigenstates. It was Gutzwiller [1], who first suggested to study three-dimensional hyperbolic billiards, being embedded in a space of constant negative curvature, as ideal models for quantum chaotic systems in higher dimensions.

Jacobi's equation for the geodesic deviation shows that in the case of constant negative curvature neighbouring geodesics always diverge at an exponential rate thus yielding one of the two ingredients of chaotic systems. The other requirement is a finite phase space which is attained by the finite configuration space of the billiard. Before we concentrate on the three-dimensional case we would like to recall some facts about the two-dimensional case. In the two-dimensional hyperbolic space with constant curvature, polygons are cut out from the hyperbolic plane such that their geodesic edges can be identified in pairs by imposing periodic boundary conditions. This can also be viewed as a tessellation of the whole hyperbolic plane. In this way a fundamental cell is obtained on which the dynamics of a free particle can be investigated. The topology of the surface is determined by the orientation and Euler's invariant $w = v - e + f$, where $v$, $e$ and $f$ denote the number of vertices, edges and faces, respectively. For a given $w$ a Teichmüller space exists which parameterizes surfaces having the same topology. The same procedure in three-dimensional hyperbolic space, i. e., cutting out polyhedra, does not yield a parametric family of spaces with the same topology. One instead obtains a variety of different topological species which cannot be described by a simple measure like Euler's invariant in two dimensions. This difference is caused by the many conditions one has to fulfill in order to get a fundamental cell which tessellates the three-dimensional hyperbolic space. To tessellate the hyperbolic space with a polyhedron one has to require that the dihedral angles around an edge add up to $2\pi$ and the solid angles forming a vertex add up to $4\pi$, as will be explained later. In addition to these conditions one has to obtain a polyhedron whose faces can be identified in pairs. These restrictions are the reason for having only "discrete" models in the three-dimensional case. To choose a simple model for the study of quantum chaos we concentrate on a pentahedron being symmetric along an intersection plane, see figure 1. (To identify faces in pairs view it as a hexahedron with one dihedral angle of $\pi$.) Desymmetrization of the pentahedron leads to a tetrahedron, the most fundamental three-dimensional object. Instead of the periodic boundary conditions of the pentahedron one has Dirichlet and Neumann boundary conditions in the desymmetrized system, which facilitates the numerical computation of quantal levels. The tetrahedron itself shows a rotational symmetry (see end of section II). If one is interested in spectral statistics it is important to desymmetrize the considered system *totally*. However, it is easier to illustrate the connection between geometric quantities and quantum energy levels, if we keep the geometry simple.

The quantum mechanics of a chaotic system can only be understood in connection with the classical counterpart. Geometric quantities of the classical system contain all information about the spectrum of quantal levels via an exact trace formula that takes over the role of Gutzwiller's semiclassical trace formula [2]. The trace formula can be used as a quantization condition for the energy levels as well as for a statistical description of the number variance $\Sigma^2(L)$ or the spectral rigidity $\Delta_3(L)$. An important contribution in the trace formula is due to the classical periodic orbits and the geometric quantities attached to them. These geometric quantities arise by the linearization of the motion along the periodic orbit thus describing its infinitesimal neighbourhood. From three dimensions on there exists an additional type of periodic orbit, the so-called loxodromic orbit. The neighbourhood of such a periodic orbit is rotated after one traversal, a property being impossible in two dimensions.

This paper is organized as follows. The next section provides together with the appendix a derivation of the trace formula for general co-compact lattices of three-dimensional hyperbolic space. The hyperbolic tetrahedron is a fundamental cell of such a hyperbolic lattice being generated by the reflections at the faces of the tetrahedron. Until now the trace formula was only available for lattice groups without reflections, i. e., for polyhedra with periodic boundary conditions. Section III describes in more detail the structure of hyperbolic reflection groups with special emphasis on lattices having a tetrahedral fundamental cell. An average law for the spectral staircase is derived from the trace formula and applied to the considered system. Section IV deals with the numerical computation of the periodic orbits, which are used for the calculation of quantal levels via the trace formula in section V, where the spectral staircase and the "cosine-quantization" are discussed. The results obtained by the trace formula are compared with numerical computations of the quantal levels using the boundary element method.

# II Geometry and trace formulae

There exist several models for hyperbolic space, see e. g. [3]. For our considerations we prefer the upper half space

$$\mathfrak{H}_3 = \{(x_1, x_2, x_3) \in \mathbb{R}^3 \mid x_3 > 0\}, \tag{1}$$

equipped with the Riemannian metric

$$ds^2 = \kappa \, \frac{dx_1^2 + dx_2^2 + dx_3^2}{x_3^2}. \tag{2}$$

The curvature equals $-1/\kappa$ at every point of $\mathfrak{H}_3$; we set $\kappa = 1$. The main advantage of this model is that the action of the group of isometries on $\mathfrak{H}_3$ (i.e., the distance preserving bijections on $\mathfrak{H}_3$; we will denote them by Iso $\mathfrak{H}_3$) has a simple representation by linear fractional transformations: Set $x = x_1 + x_2\mathrm{i} + x_3\mathrm{j}$, with the quaternions i and j defined by the relations $\mathrm{i}^2 = \mathrm{j}^2 = -1$ and $\mathrm{ij} + \mathrm{ji} = 0$, plus the property that i and j commute with every real number. The inverse of a quaternion $q = q_1 + q_2\mathrm{i} + q_3\mathrm{j} + q_4\mathrm{ij} \neq 0$ is then given by $q^{-1} = |q|^{-2}(q_1 - q_2\mathrm{i} - q_3\mathrm{j} - q_4\mathrm{ij})$, where $|q|^2 = q_1^2 + q_2^2 + q_3^2 + q_4^2$. It can be shown that every orientation preserving isometry $f$ of $\mathfrak{H}_3$ has a representation

$$f(x) = (ax + b)(cx + d)^{-1} \text{ , where } \begin{pmatrix} a & b \\ c & d \end{pmatrix} \in \mathrm{SL}(2, \mathbb{C}), \tag{3}$$

while for the orientation reversing case this matrix should be contained in the coset $\mathrm{SL}(2, \mathbb{C})\mathrm{j}$, i. e., the matrix elements are of the form $z\mathrm{j}$, $z \in \mathbb{C}$. It can easily be deduced that the group of orientation preserving isometries $\mathrm{Iso}^+\mathfrak{H}_3$ is isomorphic to $\mathrm{PSL}(2,\mathbb{C}) = \mathrm{SL}(2,\mathbb{C})/\{\pm\mathbf{1}\}$ and the class of orientation reversing isometries is isomorphic to $\mathrm{PSL}(2,\mathbb{C})\mathrm{j}$. We say that an isometry $f$ *belongs* to a matrix $F \in \mathrm{SL}(2,\mathbb{C}) \cup \mathrm{SL}(2,\mathbb{C})\mathrm{j}$.

The isometries of $\mathfrak{H}_3$ can be classified the following way [4]:

| $f$ is called ... , | if $F$ is $\mathrm{SL}(2,\mathbb{C})$-conjugate to ... |
|---|---|
| *plane reflection* | $\pm\begin{pmatrix} 1 & 0 \\ 0 & 1 \end{pmatrix}\mathrm{j}$ |
| *elliptic* | $\pm\begin{pmatrix} \mathrm{e}^{\mathrm{i}\phi/2} & 0 \\ 0 & \mathrm{e}^{-\mathrm{i}\phi/2} \end{pmatrix}, \quad \phi \in (0,\pi]$ |
| *inverse elliptic* | $\pm\begin{pmatrix} 0 & \mathrm{i}\mathrm{e}^{\mathrm{i}\phi/2} \\ \mathrm{i}\mathrm{e}^{-\mathrm{i}\phi/2} & 0 \end{pmatrix}\mathrm{j}, \quad \phi \in (0,\pi]$ |
| *parabolic* | $\pm\begin{pmatrix} 1 & 1 \\ 0 & 1 \end{pmatrix}$ |
| *inverse parabolic* | $\pm\begin{pmatrix} 1 & 1 \\ 0 & 1 \end{pmatrix}\mathrm{j}$ |
| *hyperbolic* | $\pm\begin{pmatrix} \mathrm{e}^{l/2+\mathrm{i}\phi/2} & 0 \\ 0 & \mathrm{e}^{-l/2-\mathrm{i}\phi/2} \end{pmatrix}, \quad l > 0,\ \phi \in [0,2\pi)$ |
| *inverse hyperbolic* | $\pm\begin{pmatrix} \mathrm{e}^{l/2} & 0 \\ 0 & \mathrm{e}^{-l/2} \end{pmatrix}\mathrm{j}, \quad l > 0$ |

We call $l = l_f$ the *length* and $\phi = \phi_f$ the *phase* of the transformation. In case of elliptic elements the phase $\phi$ corresponds to the rotation angle. Hyperbolic elements are called *loxodromic*, if $\phi \neq 0$.

Take a discrete subgroup $\Gamma$ of $\mathrm{Iso}\,\mathfrak{H}_3$, and identify all points of $\mathfrak{H}_3$ which can be transformed into each other by an element of $\Gamma$. Those points are called $\Gamma$-*equivalent* and we put them into an equivalence class $\Gamma(x) = \{g(x) \mid g \in \Gamma\}$ with $x \in \mathfrak{H}_3$. The set of those classes is the hyperbolic three-orbifold $\mathfrak{H}_3/\Gamma = \{\Gamma(x) \mid x \in \mathfrak{H}_3\}$.

To visualize the orbifold we take one representative from each class such that the set of all representatives yields a simply connected set in $\mathfrak{H}_3$, called *fundamental cell* $\mathcal{F}_\Gamma$. If the fundamental cell is of finite volume, we call $\Gamma$ a hyperbolic *crystallographic group* or simply a hyperbolic *lattice*. If the closure of the fundamental cell is compact, $\Gamma$ is called a co-compact lattice. A hyperbolic lattice is co-compact if and only if $\Gamma$ contains no parabolic element – otherwise the fixpoint of the parabolic transformation, which is at infinity, would be contained in the fundamental cell. Discrete subgroups of $\mathrm{Iso}^+\mathfrak{H}_3$ are called *Kleinian groups*.

The dynamics of a classical point particle moving freely in the space $\mathfrak{H}_3/\Gamma$ is strongly chaotic, if the volume of $\mathfrak{H}_3/\Gamma$ is finite. The dynamics reflect the properties of an Anosov system [5], to be precise. (This is true for any closed hyperbolic orbifold in any dimension.)

Turning to quantum mechanics we study the stationary Schrödinger equation (in units $\hbar = 2m = 1$)

$$-\Delta\psi(x) = E\psi(x), \qquad \psi \in \mathrm{L}^2(\mathfrak{H}_3/\Gamma,\chi), \tag{4}$$

$\Delta$ denotes the Laplace-Beltrami operator

$$\Delta = x_3^2\left(\frac{\partial^2}{\partial x_1^2}+\frac{\partial^2}{\partial x_2^2}+\frac{\partial^2}{\partial x_3^2}\right) - x_3\frac{\partial}{\partial x_3}. \tag{5}$$

$L^2(\mathfrak{H}_3/\Gamma,\chi)$ is the space of all functions which are square integrable on the fundamental cell and $\Gamma$-automorphic, i.e., satisfy for all $g\in\Gamma$

$$\psi(g(x)) = \chi(g)\,\psi(x), \tag{6}$$

where $\chi$ is any one-dimensional unitary representation of $\Gamma$, so-called *character*. For example take a reflection group – like we will do later on – generated by the reflections $g_1,\ldots,g_n$ at faces of a hyperbolic polyhedron, and choose $\chi(g_i)=-1$ for all those generators, then one will obtain Dirichlet boundary conditions on the faces of the polyhedron. Choosing $\chi(g_i)=1$ yields Neumann boundary conditions.

In the following we will study the eigenvalues of the operator $-\Delta$ on a compact orbifold $\mathfrak{H}_3/\Gamma$, where the spectrum is discrete,

$$0\le E_1\le E_2\le\cdots.$$

Difficulties arising in the non-compact case from the continuous part of the spectrum are of pure technical nature. They will cause some additional terms in the trace formula which are of minor importance. Therefore let us not lose track and concentrate on compact orbifolds. To obtain information about the energy spectrum we look at the trace of the resolvent of equation (4),

$$\operatorname{tr}\left[(-\Delta-E)^{-1}\right] = \sum_{n=1}^{\infty}(E_n-E)^{-1}. \tag{7}$$

This series diverges, however, as the number of eigenvalues below an energy $E$ is asymptotically ($E\to\infty$) proportional to $E^{3/2}$ according to Weyl's law, so the energies $E_n$ grow just like $n^{2/3}$. Let us replace the divergent trace by

$$\operatorname{tr}\left[(-\Delta-E)^{-1}-(-\Delta-E')^{-1}\right] = \sum_{n=1}^{\infty}\left[(E_n-E)^{-1}-(E_n-E')^{-1}\right], \tag{8}$$

which is absolutely convergent. Since $E'\neq E_n$ is fixed, the poles of the regularized resolvent are still at the eigenvalues. The integral kernel of the resolvent is the Green function $G_\Gamma(x,x';E)$ of equation (4) which is a coherent superposition of the free Green function $G(x,x';E)$ on $\mathfrak{H}_3$,

$$G_\Gamma(x,x';E) = \sum_{g\in\Gamma}\chi(g)\,G(g(x),x';E), \tag{9}$$

where

$$G(x,x';E) = \frac{1}{4\pi\sinh d(x,x')}\,\exp[\mathrm{i}\,p\,d(x,x')], \qquad p=\sqrt{E-1}. \tag{10}$$

$d(x,x')$ denotes the hyperbolic distance between $x$ and $x'\in\mathfrak{H}_3$, defined by

$$\cosh d(x,x') = 1+\frac{|x-x'|^2}{2x_3x_3'}. \tag{11}$$

It should be noted that the three-dimensional Green function contains only elementary functions, while in two dimensions Legendre functions are involved – a characteristic difference between even and odd dimension.

Finally we are left with the integral

$$\text{tr}\left[(-\Delta - E)^{-1} - (-\Delta - E')^{-1}\right] = \int_{\mathcal{F}_\Gamma} d\mu(x)\, [G_\Gamma(x,x;E) - G_\Gamma(x,x;E')], \tag{12}$$

$\mu(x)$ being the volume element of $\mathfrak{H}_3$,

$$d\mu(x) = \frac{dx_1 dx_2 dx_3}{x_3^3}. \tag{13}$$

The integration is performed in the appendix, the result reads

$$\begin{aligned}
&\sum_{n=1}^{\infty} \left[(p_n^2 - p^2)^{-1} - (p_n^2 - p'^2)^{-1}\right] \\
&= -\frac{\text{Vol}(\mathcal{F}_\Gamma)}{4\pi\mathrm{i}}(p - p') \\
&\quad - \sum_{\substack{\{\rho\}_\Gamma \\ \text{inv.}}} \chi(\rho)\,\frac{\text{Area}(\mathcal{P}_\rho)}{16\pi}\,[\psi(1-\mathrm{i}p) + \psi(-\mathrm{i}p) - \psi(1-\mathrm{i}p') - \psi(-\mathrm{i}p')] \\
&\quad - \sum_{\substack{\{\rho\}_\Gamma \\ \text{ellipt.}}} \frac{\chi(\rho)\, l_{\tau_0}}{4\,\text{ord}\mathcal{E}_\Gamma(\rho)\,(1-\cos\phi_\rho)} \left(\frac{1}{\mathrm{i}p} - \frac{1}{\mathrm{i}p'}\right) \\
&\quad + \sum_{\substack{\{\rho\}_\Gamma \\ \text{inv. ellipt.}}} \frac{\chi(\rho)}{\text{ord}\mathcal{E}_\Gamma(\rho)}[I(p,\phi_\rho) - I(p',\phi_\rho)] \\
&\quad - \sum_{\substack{\{\tau\}_\Gamma \\ \text{hyperbol.}}} \frac{\chi(\tau)\, l_{\tau_0}}{4\,\text{ord}\mathcal{E}_\Gamma(\tau)\,(\cosh l_\tau - \cos\phi_\tau)} \left(\frac{\exp(\mathrm{i}\,p\,l_\tau)}{\mathrm{i}p} - \frac{\exp(\mathrm{i}\,p'\,l_\tau)}{\mathrm{i}p'}\right) \\
&\quad - \sum_{\substack{\{\tau\}_\Gamma \\ \text{inv. hyperbol.}}} \frac{\chi(\tau)\, l_{\tau_0}}{4\,\text{ord}\mathcal{E}_\Gamma(\tau)\,\sinh l_\tau} \left(\frac{\exp(\mathrm{i}\,p\,l_\tau)}{\mathrm{i}p} - \frac{\exp(\mathrm{i}\,p'\,l_\tau)}{\mathrm{i}p'}\right),
\end{aligned} \tag{14}$$

for $\text{Im}\,p > 1$, $\text{Im}\,p' > 1$. We have substituted the energy variables by momentum variables, setting $E = p^2 + 1$, $E' = p'^2 + 1$ and $E_n = p_n^2 + 1$, where $-\mathrm{i}p_n \in (0,1]$ for $n = 1, \dots, N$, if $N$ is the number of all so-called "small" eigenvalues $E_n \in [0,1)$, and $p_n \geq 0$ for $n > N$.

The sums in (14) extend over $\Gamma$-conjugacy classes

$$\{g\}_\Gamma \;:=\; \{\, g' \,|\, g' = h\, g\, h^{-1} \,,\; h \in \Gamma \,\} \tag{15}$$

of plane reflections, elliptic, inverse elliptic, hyperbolic and inverse hyperbolic transformations, denoted by $\rho$ or $\tau$, respectively. $\mathcal{P}_\rho$ is a collection of all parts of the fundamental cell, which are left invariant by the reflection $\rho$ or one of its conjugates. $\tau_0$ is a transformation with a shortest length $l_{\tau_0}$ of all hyperbolic and inverse hyperbolic transformations commuting with $\rho$ or $\tau$, respectively. $\text{ord}\mathcal{E}_\Gamma(f)$ denotes the order of a maximal finite subgroup $\mathcal{E}_\Gamma(f)$ of the centralizer $\mathcal{C}_\Gamma(f)$, which is the subgroup of $\Gamma$ that contains all elements commuting with $f$.

$\psi(x)$ is the logarithmic derivative of Euler's gamma function, the function $I(p,\phi)$ is defined in the appendix.

The distinct contributions to the trace formula can be interpreted in a nice geometric way: The first summand depends on the volume of $\mathfrak{H}_3/\Gamma$. The second contribution is connected with the orbifold's surface (if there is any), it is a sum over distinct surface parts belonging to chosen boundary conditions. If one chooses a unique boundary condition on the whole surface (e.g. Neumann or Dirichlet type), one will obtain a contribution

$$\mp\frac{\text{surface area}}{16\pi}\,[\psi(1-\mathrm{i}p)+\psi(-\mathrm{i}p)-\psi(1-\mathrm{i}p')-\psi(-\mathrm{i}p')],$$

upper sign for Neumann, lower sign for Dirichlet boundary conditions. The third contribution is connected with lengths and dihedral angles of the edges of $\mathfrak{H}_3/\Gamma$, the fourth one with its corner angles, in a way that will be explained by example in section III. There is a correspondence between closed geodesics of $\mathfrak{H}_3/\Gamma$ and conjugacy classes of hyperbolic and inverse hyperbolic elements of $\Gamma$. This relation is, however, in general not one-to-one: Let $\{\tau\}_\Gamma$ be a hyperbolic or inverse hyperbolic conjugacy class associated with a periodic orbit. Then all conjugacy classes of the form $\{\tau\rho\}_\Gamma$, $\rho\in\mathcal{E}_\Gamma(\tau)$ belong to the same orbit. However, $\mathrm{ord}\mathcal{E}_\Gamma(\tau)\neq 1$ only for surface or edge orbits. Periodic orbits for which $\mathrm{ord}\mathcal{E}_\Gamma(\tau)=1$ will be called *interior* orbits. $l_\tau$ corresponds to the length of the periodic orbit associated with $\{\tau\}_\Gamma$.

Let us compare our periodic-orbit contribution to the trace of the resolvent with that one obtained by Gutzwiller's theory [2], which is given by

$$-\frac{\chi_{\mathrm{PO}}\,l_{\mathrm{PPO}}}{2\,|\det(M_{\mathrm{PO}}-\mathbf{1})|^{1/2}}\left(\frac{\exp(\mathrm{i}\,p\,l_{\mathrm{PO}})}{\mathrm{i}p}-\frac{\exp(\mathrm{i}\,p'\,l_{\mathrm{PO}})}{\mathrm{i}p'}\right), \tag{16}$$

where $\chi_{\mathrm{PO}}$ is a phase factor involving the Morse index of the considered periodic orbit PO and boundary conditions, $l_{\mathrm{PO}}$ its length, and $l_{\mathrm{PPO}}$ the length of the associated primitive periodic orbit PPO. The monodromy matrix $M_{\mathrm{PO}}$ is a linearized map of the four-dimensional surface of section (at an arbitrary point $P$ on the periodic orbit) onto itself, which describes the traversal of phase space trajectories, starting and ending at points on the surface of section nearby $P$.

With every interior orbit we can associate exactly one conjugacy class $\{\tau\}_\Gamma$, where $\tau$ is hyperbolic or inverse hyperbolic, depending on the periodic orbit being direct or inverse hyperbolic. The monodromy matrix is a map $M_{\mathrm{PO}}:\mathbb{R}^4\to\mathbb{R}^4$, $(\delta q_1,\delta q_2,\delta p_1,\delta p_2)^{\mathrm{t}}\mapsto M_{\mathrm{PO}}(\delta q_1,\delta q_2,\delta p_1,\delta p_2)^{\mathrm{t}}$. $\delta q_i$ are local position coordinates and $\delta p_i$ local momentum coordinates of a coordinate system orthogonal to the periodic orbit at $P$. A straightforward calculation shows that

$$|\det(M_{\mathrm{PO}}-\mathbf{1})|^{1/2}=\begin{cases}2\,(\cosh l_{\mathrm{PO}}-\cos\phi_{\mathrm{PO}}) & \text{if the periodic orbit is direct hyperbolic}\\ 2\,\sinh l_{\mathrm{PO}} & \text{if the periodic orbit is inverse hyperbolic,}\end{cases} \tag{17}$$

in accordance with trace formula (14).

This result can easily be generalized to the $n$-dimensional case, where we obtain for $n$ odd

$$|\det(M_{\mathrm{PO}}-\mathbf{1})|^{1/2}=2^k\cdot\prod_{i=1}^{j}(\cosh l_{\mathrm{PO}}-\cos\phi_{i,\mathrm{PO}})\cdot(\sinh l_{\mathrm{PO}})^{k-j},$$

with $k:=(n-1)/2$, and $j$ phases $\phi_{i,\mathrm{PO}}$, $j\in\{0,\dots,k\}$. The considered periodic orbit PO is direct hyperbolic, if $k-j$ is even, and inverse hyperbolic otherwise. In the even dimensional

case we have

$$|\det(M_{\rm PO}-\mathbf{1})|^{1/2} = \begin{cases} 2^k \cdot \cosh(l_{\rm PO}/2) \cdot \prod_{i=1}^{j}(\cosh l_{\rm PO} - \cos\phi_{i,{\rm PO}}) \cdot (\sinh l_{\rm PO})^{k-j-1} \\ 2^k \cdot \sinh(l_{\rm PO}/2) \cdot \prod_{i=1}^{j}(\cosh l_{\rm PO} - \cos\phi_{i,{\rm PO}}) \cdot (\sinh l_{\rm PO})^{k-j-1}, \end{cases}$$

with $k := n/2$, and $j$ phases $\phi_{i,{\rm PO}}$, $j \in \{0,\dots,k-1\}$. The upper (resp. lower) case corresponds to a direct hyperbolic orbit, if $k-j-1$ is even (resp. odd), and to an inverse hyperbolic orbit otherwise. Let us return to $n=3$.

In the case of surface or edge orbits $M_{\rm PO}$ cannot be determined in a unique way: Let $\mathcal{E}$ be the discrete subgroup of O(2) that leaves the edge (resp. the surface part) pointwise invariant. $\mathcal{E}$ identifies points on the surface of section via the action defined by

$$\begin{pmatrix} \rho & \mathbf{0} \\ \mathbf{0} & \rho \end{pmatrix} (\delta q_1, \delta q_2, \delta p_1, \delta p_2)^{\rm t}, \qquad \rho \in \mathcal{E}, \tag{18}$$

where $\mathbf{0}$ denotes the $2\times 2$ zero matrix. Therefore $M_{\rm PO}$ is a map $M_{\rm PO} : \mathbb{R}^4/\mathcal{E} \to \mathbb{R}^4/\mathcal{E}$, and with $M_{\rm PO}$ there are additional (ord$\mathcal{E}-1$) matrices of the type

$$\tilde{M}_{\rm PO} = M_{\rm PO} \begin{pmatrix} \rho & \mathbf{0} \\ \mathbf{0} & \rho \end{pmatrix}, \qquad \rho \in \mathcal{E}, \tag{19}$$

which do not differ in their action on $\mathbb{R}^4/\mathcal{E}$, but have different determinants $\det(\tilde{M}_{\rm PO}-\mathbf{1})$ in general. Let us see what contribution instead of (16) is suggested by trace formula (14).

First it can be shown that there exists a hyperbolic element $\tau$ associated with the considered periodic orbit, such that $\mathcal{E}_\Gamma(\tau)$ is isomorphic to $\mathcal{E}$. If two elements $\rho_1$, $\rho_2$ of $\mathcal{E}_\Gamma(\tau)$ are $\mathcal{E}_\Gamma(\tau)$-conjugate, then $\{\tau\rho_1\}_\Gamma = \{\tau\rho_2\}_\Gamma$. It is an algebraic exercise to show that, if $k$ elements $\rho_1,\dots,\rho_k$ of $\mathcal{E}_\Gamma(\tau)$ are $\mathcal{E}_\Gamma(\tau)$-conjugate, then

$$\text{ord}\mathcal{E}_\Gamma(\tau\rho_i) = \text{ord}\mathcal{E}_\Gamma(\tau)/k, \qquad i=1,\dots,k. \tag{20}$$

Thus

$$\sum_{\substack{\{\tau\rho\}_\Gamma \\ \tau \text{ fixed} \\ \rho\in\mathcal{E}_\Gamma(\tau)}} \frac{1}{\text{ord}\mathcal{E}_\Gamma(\tau\rho)} = \frac{1}{\text{ord}\mathcal{E}_\Gamma(\tau)} \sum_{\rho\in\mathcal{E}_\Gamma(\tau)} 1 = 1, \tag{21}$$

where the first sum runs over all distinct conjugacy classes $\{\tau\rho\}_\Gamma$, which means that the elements $\rho \in \mathcal{E}_\Gamma(\tau)$ are not $\mathcal{E}_\Gamma(\tau)$-conjugate to each other. $\tau$ is fixed and satisfies the conditions described above. The contribution of a surface or an edge orbit to (14) is then given by

$$-\Bigg[ \sum_{\substack{\{\tau\rho\}_\Gamma \\ \tau \text{ fixed} \\ \rho\in\mathcal{E}^+_\Gamma(\tau)}} \frac{\chi(\tau\rho)\, l_{\tau_0}}{4\,\text{ord}\mathcal{E}_\Gamma(\tau\rho)\,(\cosh l_\tau - \cos\phi_{\tau\rho})} + \sum_{\substack{\{\tau\rho\}_\Gamma \\ \tau \text{ fixed} \\ \rho\in\mathcal{E}^-_\Gamma(\tau)}} \frac{\chi(\tau\rho)\, l_{\tau_0}}{4\,\text{ord}\mathcal{E}_\Gamma(\tau\rho)\, \sinh l_\tau} \Bigg] \times \\ \times \left( \frac{\exp(\mathrm{i}\, p\, l_\tau)}{\mathrm{i}p} - \frac{\exp(\mathrm{i}\, p'\, l_\tau)}{\mathrm{i}p'} \right), \tag{22}$$

or, written as a sum over all elements of $\mathcal{E}_\Gamma(\tau)$,

$$(23)\quad -\frac{l_{\tau_0}}{4\,\mathrm{ord}\mathcal{E}_\Gamma(\tau)}\Bigg[\sum_{\rho\in\mathcal{E}^+_\Gamma(\tau)}\frac{\chi(\tau\rho)}{\cosh l_\tau-\cos\phi_{\tau\rho}}+\sum_{\rho\in\mathcal{E}^-_\Gamma(\tau)}\frac{\chi(\tau\rho)}{\sinh l_\tau}\Bigg]\left(\frac{\exp(\mathrm{i}\,p\,l_\tau)}{\mathrm{i}p}-\frac{\exp(\mathrm{i}\,p'\,l_\tau)}{\mathrm{i}p'}\right),$$

where $\mathcal{E}^\pm_\Gamma(\tau)$ denotes the set of orientation preserving and orientation reversing elements of $\mathcal{E}_\Gamma(\tau)$, respectively. In the case of edge and surface orbits we should therefore replace (16) by

$$(24)\qquad -\frac{l_{\mathrm{PPO}}}{2\,\mathrm{ord}\mathcal{E}}\sum_{\rho\in\mathcal{E}}\frac{\tilde{\chi}_{\mathrm{PO}}}{|\det(\tilde{M}_{\mathrm{PO}}-\mathbf{1})|^{1/2}}\left(\frac{\exp(\mathrm{i}\,p\,l_{\mathrm{PO}})}{\mathrm{i}p}-\frac{\exp(\mathrm{i}\,p'\,l_{\mathrm{PO}})}{\mathrm{i}p'}\right),$$

which is nothing but the average over all contributions (16) with different $\rho\in\mathcal{E}$. For the definition of $\tilde{M}_{\mathrm{PO}}$ see (19), $\tilde{\chi}_{\mathrm{PO}}$ is the corresponding phase factor.

Let us put the trace formula into a slightly more general form by introducing test functions $h:\mathbb{C}\to\mathbb{C}$ with the properties

(i) $h$ is even, i.e., $h(p)=h(-p)$,

(ii) $h$ is holomorphic in the strip $|\mathrm{Im}\,p|\le\sigma$, $\sigma>1$,

(iii) $h(p)=O(|p|^{-3-\delta})$ for $\delta>0$ as $|p|\to\infty$.

We multiply (14) by $p\,h(p)/\pi\mathrm{i}$ and integrate over $p$ from $-\infty+\mathrm{i}\sigma$ to $\infty+\mathrm{i}\sigma$ to obtain the general trace formula for co-compact hyperbolic lattices in $\mathfrak{H}_3$,

$$\begin{aligned}\sum_{n=1}^{\infty}h(p_n)=&-\frac{\mathrm{Vol}(\mathcal{F}_\Gamma)}{2\pi}\,\tilde{h}''(0)\\
&+\sum_{\substack{\{\rho\}_\Gamma\\ \text{inv.}}}\chi(\rho)\,\frac{\mathrm{Area}(\mathcal{P}_\rho)}{8\pi}\int_0^\infty dq\;q\,h(q)\,\coth(\pi q)\\
&+\sum_{\substack{\{\rho\}_\Gamma\\ \text{ellipt.}}}\frac{\chi(\rho)\,l_{\tau_0}}{2\,\mathrm{ord}\mathcal{E}_\Gamma(\rho)\,(1-\cos\phi_\rho)}\,\tilde{h}(0)\\
(25)\qquad&+\sum_{\substack{\{\rho\}_\Gamma\\ \text{inv. ellipt.}\\ 0<\phi<\pi}}\frac{\chi(\rho)}{2\,\mathrm{ord}\mathcal{E}_\Gamma(\rho)}\int_0^\infty dq\;h(q)\,\frac{\sinh[(\pi-\phi_\rho)q]}{\sinh(\pi q)\;\sin\phi_\rho}\\
&+\sum_{\substack{\{\rho\}_\Gamma\\ \text{inv. ellipt.}\\ \phi=\pi}}\frac{\chi(\rho)}{2\,\mathrm{ord}\mathcal{E}_\Gamma(\rho)}\int_0^\infty dq\;h(q)\,\frac{q}{\sinh(\pi q)}\\
&+\sum_{\substack{\{\tau\}_\Gamma\\ \text{hyperbol.}}}\frac{\chi(\tau)\,l_{\tau_0}}{2\,\mathrm{ord}\mathcal{E}_\Gamma(\tau)\,(\cosh l_\tau-\cos\phi_\tau)}\,\tilde{h}(l_\tau)\\
&+\sum_{\substack{\{\tau\}_\Gamma\\ \text{inv. hyperbol.}}}\frac{\chi(\tau)\,l_{\tau_0}}{2\,\mathrm{ord}\mathcal{E}_\Gamma(\tau)\,\sinh l_\tau}\,\tilde{h}(l_\tau)\;.\end{aligned}$$

$\tilde{h}$ is the Fourier transform of $h$:

$$\tilde{h}(x) = \frac{1}{2\pi} \int\limits_{-\infty}^{\infty} dp\; h(p) \exp(\mathrm{i}px),$$

and $\tilde{h}''$ is its second derivative.

We will illustrate the trace formula in case of a lattice $\Gamma^+$ whose fundamental cell is the pentahedron shown in figure 1. (The index + will become clear below.) $\Gamma^+$ is generated by three rotations: a half turn $\rho_{BC}$ through the axis $BC$ identifying face $A'BC$ with $ABC$, a $\frac{2\pi}{3}$-turn through $DB$ identifying face $DA'B$ with $DAB$, and a $\frac{2\pi}{5}$-turn through $DC$ identifying face $CDA'$ with $CDA$. The eigenfunctions $\psi(x)$ of the Hamiltonian $-\Delta$ on $\mathrm{L}^2(\mathfrak{H}_3/\Gamma^+, 1)$ satisfy the periodicity conditions

$$\psi(g(x)) = \psi(x), \tag{26}$$

for all $g \in \Gamma^+$. As the pentahedron is symmetric with respect to the $BCD$ plane, the energy spectrum decomposes into two subspectra, belonging to eigenfunctions that are symmetric and anti-symmetric with respect to a reflection $g_A$ at $BCD$:

$$\psi(g_A(x)) = \pm\psi(x). \tag{27}$$

Condition (26) and (27) can be combined to

$$\psi(g(x)) = \chi(g)\psi(x), \tag{28}$$

for all $g$ in the group $\Gamma = \Gamma^+ \cup \Gamma^+ g_A$, where $\chi(g) = 1$, if $g$ is contained in $\Gamma^+$, and $\chi(g) = \pm 1$, if $g$ is in $\Gamma^+ g_A$. The fundamental cell of the lattice $\Gamma$ is the tetrahedron $T_8 = ABCD$. Equation (28) induces Neumann or Dirichlet boundary conditions, respectively.

It should be noted that the tetrahedron $T_8$ is symmetric with respect to a half turn $\rho_\pi$ with axis running through the midpoints of $AB$ and $CD$. Therefore again our energy spectrum decomposes into two further symmetry classes. Anyway, as the main target of this work is to demonstrate the connection between geometry and energy spectra, we would like to keep the geometry as simple as possible. The total decomposition of our spectrum will be postponed to a later publication, where we focus on spectral properties.

The next section provides some facts about lattices generated by reflections, especially tetrahedral lattices.

# III Hyperbolic Tetrahedra

Let $\Gamma$ be a reflection group, i.e., a lattice generated by the reflections $g_1, \ldots, g_n$ at the faces of a polyhedron. The discreteness of $\Gamma$ implies the following conditions on the $g_i$ [6]: Firstly each pair of generators should generate a discrete subgroup $\langle g_i, g_j \rangle$ of O(2). Remember that the product $g_i g_j$ will be a rotation through an angle $2\pi/k$, $k \in \mathbb{N}$, that is twice the dihedral angle at which both mirror planes meet. The Coxeter diagram of such a simplex looks like

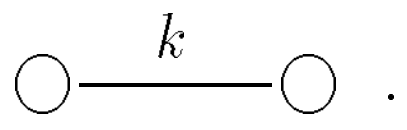

A vertex ◯ describes a plane mirror, the graph an edge of order $k$. For $k = 2, 3, 4, \ldots$ we draw alternatively

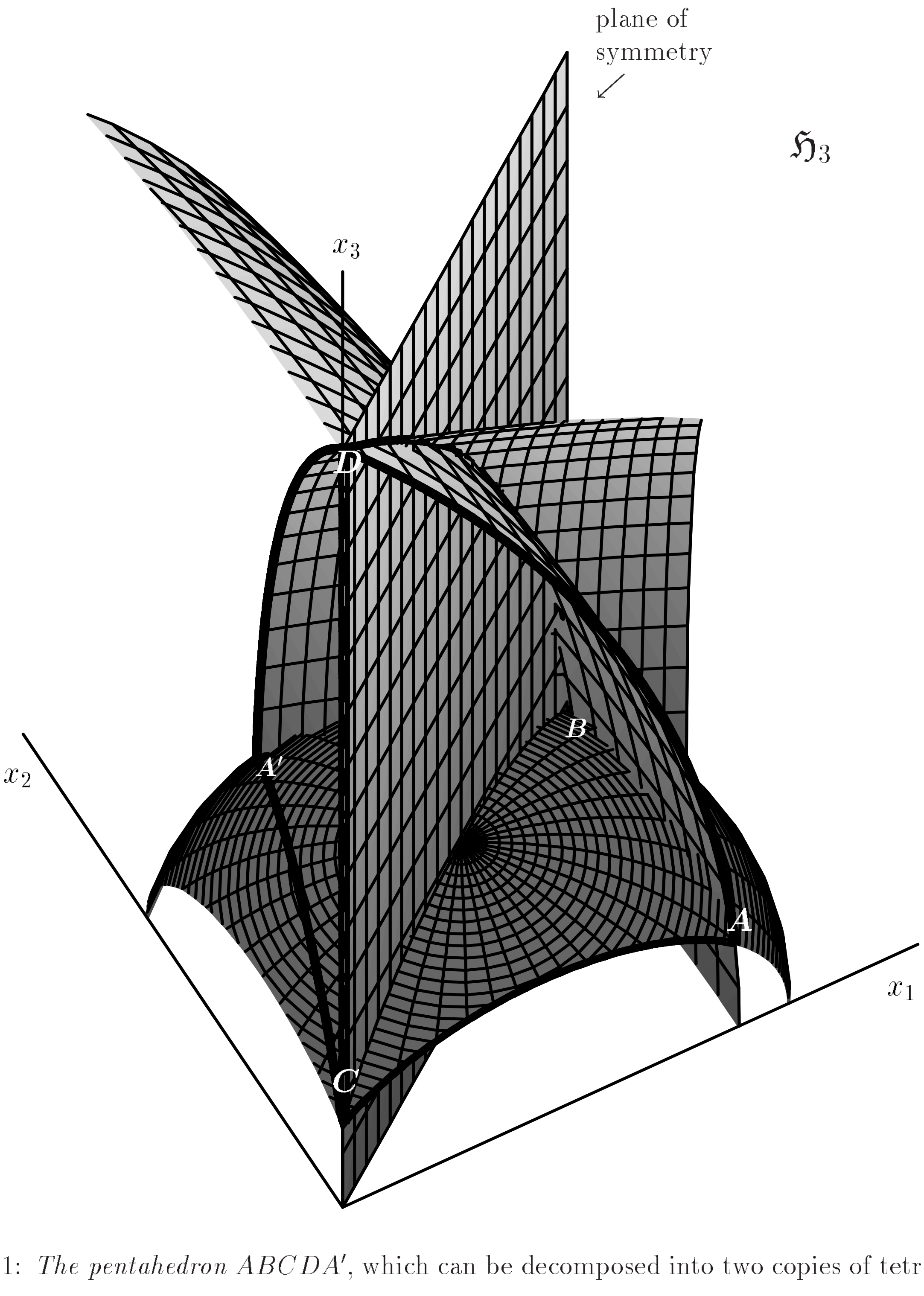

Figure 1: *The pentahedron* $ABCDA'$, which can be decomposed into two copies of tetrahedron $T_8$.

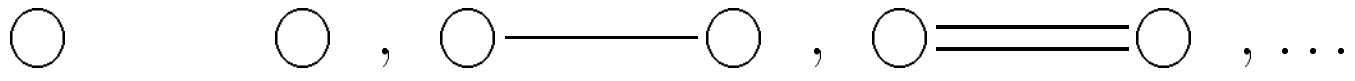

Let us turn to the corner points, at which at least three mirrors meet. The subgroup of $\Gamma$ that leaves a corner point invariant has to be a discrete subgroup of O(3). Imagine the subgroup acts on the two-sphere, then its fundamental cell should be a spherical polygon, whose corner angles correspond to the dihedral angles between the mirrors. For example suppose a corner surrounded by three mirrors, you will obtain a spherical triangle. The sum of angles in a spherical triangle is greater than $\pi$, and we get the condition

$$\frac{1}{k}+\frac{1}{l}+\frac{1}{m}>1, \qquad k,l,m\in\mathbb{N}-\{1\}, \tag{29}$$

i.e., we can admit corners of the type

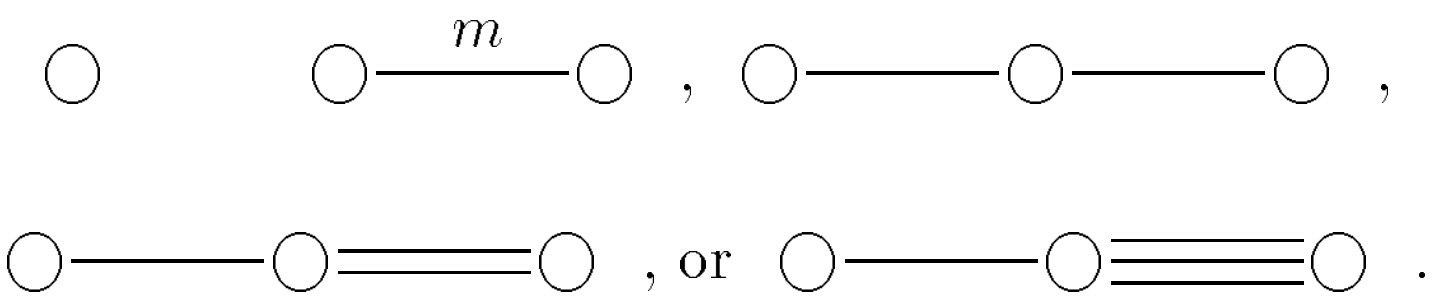

Combine these corners to, e.g., a tetrahedron, which can be embedded into spherical, Euclidean or hyperbolic space. There are exactly nine compact hyperbolic tetrahedra $T_i$ [7], out of which only $T_8$ is not arithmetic [8]. In addition we can construct non-compact tetrahedra having a *cusp at infinity*, instead of a corner. For a cusp condition (29) should be replaced by

$$\frac{1}{k}+\frac{1}{l}+\frac{1}{m}=1.$$

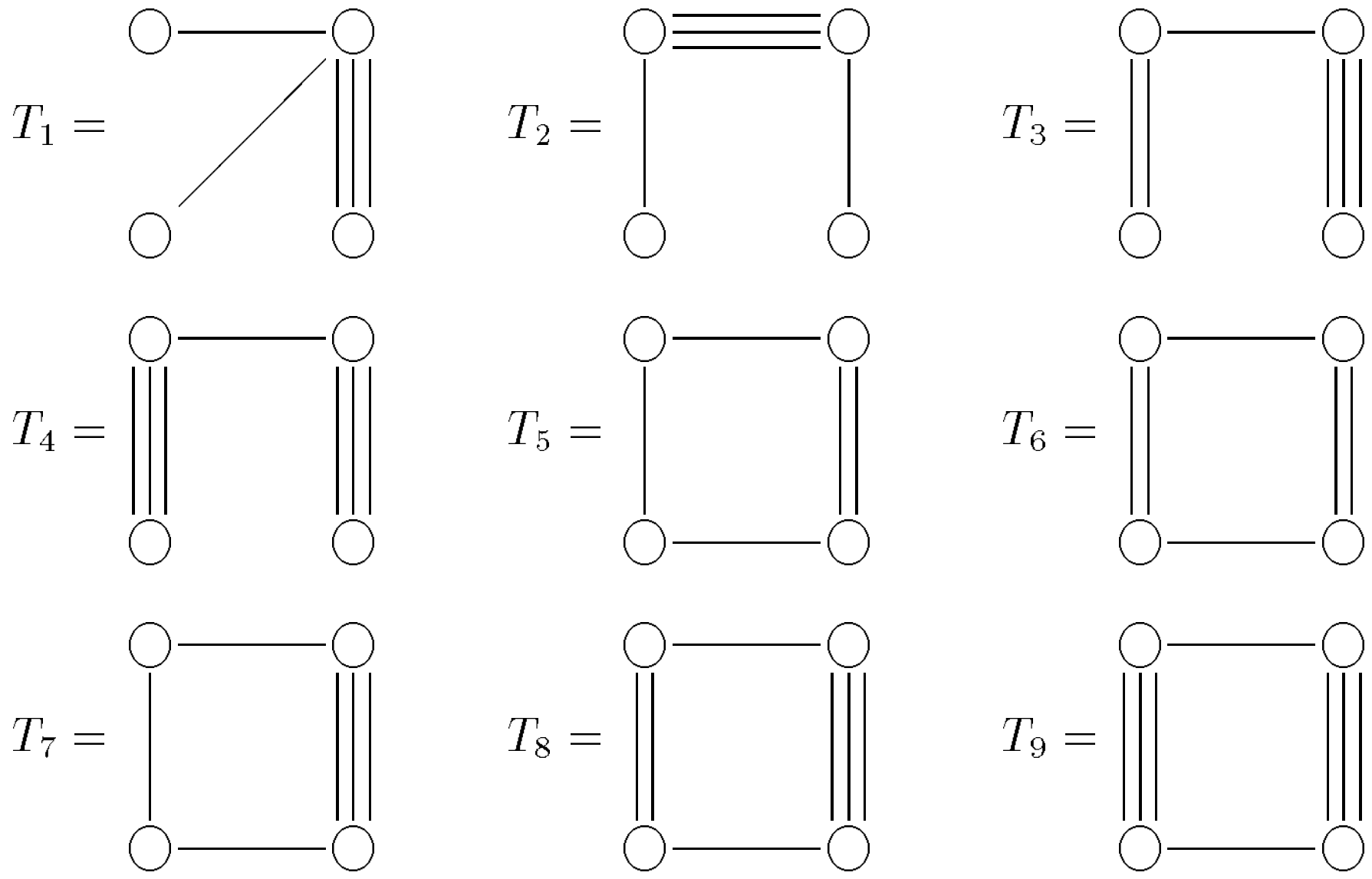

Anyway, let us stick to the compact case, and have a closer look at the tetrahedron $T_8$ as well as its reflection group, which we will denote by $\Gamma$.

Tetrahedron $T_8$ is shown in figure 1 with corner points $A,B,C,D$ and dihedral angles

$$\angle BC=\pi/2, \qquad \angle CA=\pi/3, \qquad \angle AB=\pi/4,$$

$$\angle DA = \pi/2, \qquad \angle DB = \pi/3, \qquad \angle DC = \pi/5.$$

The volume of $T_8$ is $\mathrm{Vol}(T_8) \approx 0.358653$. We cut open the tetrahedron at the even edges $BC$, $AB$ and $DA$, and unfold the surface into the $BCD$-plane. We then obtain a hexagon (see figure 2) with corner angles

$$\pi/2,\ \pi/2,\ \pi/4,\ \pi/2,\ \pi/2,\ \pi/4,$$

which is a fundamental cell of a two-dimensional lattice, generated by the reflections (see [9])

$$(30) \qquad \begin{cases} h_1 = g_C g_A g_D \; g_C \; (g_C g_A g_D)^{-1} \\ h_2 = g_D \\ h_3 = (g_B g_A)^2 g_D g_B \; g_A \, [(g_B g_A)^2 g_D g_B]^{-1} \\ h_4 = (g_B g_A)^2 g_D g_B g_C \; g_D \, [(g_B g_A)^2 g_D g_B g_C]^{-1} \\ h_5 = (g_B g_A)^2 \; g_C \; (g_B g_A)^{-2} \\ h_6 = g_C g_A \; g_B \; (g_C g_A)^{-1}, \end{cases}$$

where $g_A, g_B, g_C, g_D$ are the reflections at the face opposite to point $A, B, C, D$, respectively.

Following the Gauß-Bonnet theorem, the surface area equals $3\pi/2$. As each face of $T_8$ can be rotated into the $BCD$-plane by an element of $\Gamma$, each reflection $g_B, g_C, g_D$ is $\Gamma$-conjugate to $g_A$. We have only one conjugacy class of plane reflections, hence one can only choose identical boundary conditions on all faces of the surface, either Neumann or Dirichlet type.

Next consider elliptic conjugacy classes. First of all, each rotation of $\Gamma$ is conjugate to its inverse, as

$$(31) \qquad g_i^{-1}(g_i g_j) g_i = g_j g_i = (g_i g_j)^{-1}, \qquad \text{for } i, j = A, B, C, D.$$

Rotations through distinct angles cannot be conjugate. Let us check if the half turns through $BC$ and $DA$ are $\Gamma$-conjugate, or – which is equivalent – if the axes $BC$ and $DA$ are $\Gamma$-equivalent. Suppose they are, then there must be an element $g \in \Gamma$ transforming $B$ into $A$ and $C$ into $D$. $g$ maps $A$ and $D$ onto points that are equivalent to $B$ and $C$, respectively. Therefore $g$ is $\Gamma$-conjugate to the symmetry $\rho_\pi$ exchanging $A$ with $B$ and $C$ with $D$, which is not contained in $\Gamma$. Hence $g \notin \Gamma$ – contradiction! Similar arguments show the non-equivalence of the turns through $CA$ and $DB$. Table 1 gives an overview over all elliptic conjugacy classes. It should be noted that for $\phi_\rho \neq \pi$ we have $\mathrm{ord}\mathcal{E}_\Gamma = 4\pi/\phi_{\rho_0}$, where $\rho_0$ denotes the smallest rotation such that $\rho = \rho_0^k$, with a suitable $k \in \mathbb{N}$. For $\phi_\rho = \pi$, we have $\mathrm{ord}\mathcal{E}_\Gamma = 8\pi/\phi_{\rho_0}$, if there is another half turn in $\Gamma$ with axis orthogonal to the axis of $\rho$. If not, then $\mathrm{ord}\mathcal{E}_\Gamma = 4\pi/\phi_{\rho_0}$. In our case we get $\mathrm{ord}\mathcal{E}_\Gamma = 8\pi/\phi_{\rho_0}$ for all half turns, just have a look at the axes of the corresponding half turns in the hexagon.

Let us turn to inverse elliptic conjugacy classes. The corner points are not $\Gamma$-equivalent to one another, otherwise there would be a transformation conjugate to the symmetry $\rho_\pi$ (similar arguments as above). Therefore inverse elliptic elements leaving distinct corner points invariant are not conjugate.

The dihedral angles at a corner of our tetrahedron $T_8$ are of the form $(\pi/2, \pi/3, \pi/n)$, where $n = 4$ for corners $A$ and $B$, $n = 5$ for corners $C$ and $D$. Let $\mathcal{E}_n$ be the subgroup of all elements of $\Gamma$ leaving a particular corner point invariant, it is generated by

$$(32) \qquad \mathcal{E}_n = \langle g_i, g_j, g_k \mid (g_i g_j)^2 = (g_j g_k)^3 = (g_k g_i)^n = \mathbf{1} \rangle.$$

The order of $\mathcal{E}_4$ equals 48, the order of $\mathcal{E}_5$ equals 120.

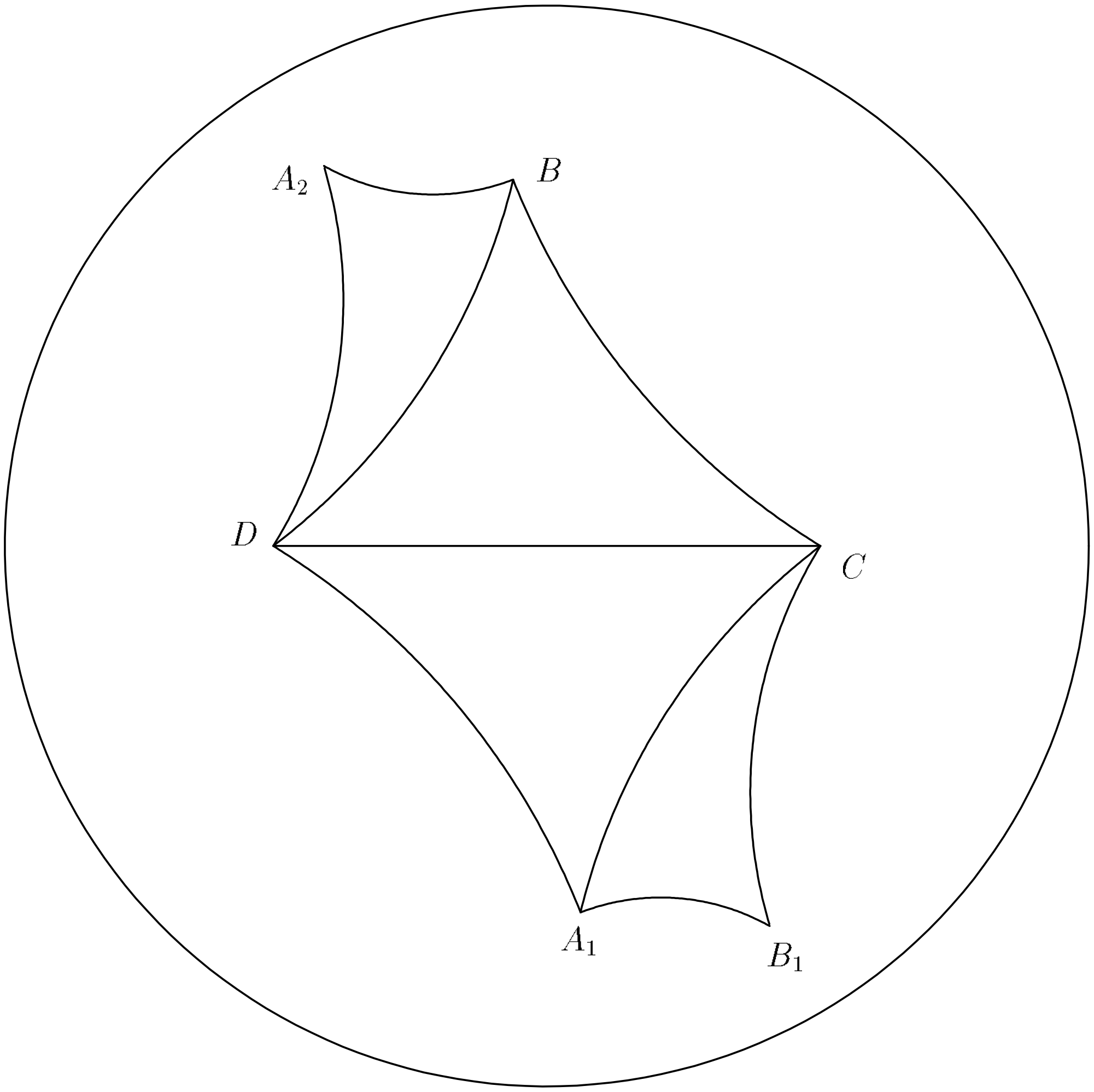

Figure 2: *Unfolded surface of tetrahedron* $T_8$. The resulting hexagon with corner angles $\pi/2, \pi/2, \pi/4, \pi/2, \pi/2, \pi/4$ is presented in the Poincaré disc $\mathfrak{D} = \{z \in \mathbb{C} \mid |z| < 1\}$, with metric $ds^2 = 4(1-|z|^2)^{-2}(dz\,d\overline{z})$. In this model of two-dimensional hyperbolic space, the geodesics are straight lines and circles perpendicular to the unit circle.

| axis | representative $\rho$ | phase $\phi_\rho$ | length $l_{\tau_0}$ | $\mathrm{ord}\mathcal{E}_\Gamma(\rho)$ |
|---|---|---|---|---|
| $BC$ | $g_D g_A$ | $\pi$ | $2d(B,C)$ | 8 |
| $CA$ | $g_D g_B$ | $2\pi/3$ | $2d(C,A)$ | 6 |
| $AB$ | $g_D g_C$ | $\pi/2$ | $2d(A,B)$ | 8 |
| $AB$ | $(g_D g_C)^2$ | $\pi$ | $2d(A,B)$ | 16 |
| $DA$ | $g_B g_C$ | $\pi$ | $2d(D,A)$ | 8 |
| $DB$ | $g_C g_A$ | $2\pi/3$ | $2d(D,B)$ | 6 |
| $DC$ | $g_A g_B$ | $2\pi/5$ | $2d(D,C)$ | 10 |
| $DC$ | $(g_A g_B)^2$ | $4\pi/5$ | $2d(D,C)$ | 10 |

Table 1: Elliptic conjugacy classes. The lengths of edges can be calculated with the second laws of cosines of hyperbolic and spherical geometry. We have $d(B,C) = d(D,A) \approx 2.273112$, $d(C,A) = d(D,B) \approx 2.132751$, $d(A,B) \approx 1.487102$, and $d(D,C) \approx 2.224036$.

| corner | representative $\rho$ | phase $\phi_\rho$ | ord$\mathcal{E}_\Gamma(\rho)$ |
|---|---|---|---|
| $A$ | $g_B g_C g_D$ | $\pi/3$ | 6 |
| $A$ | $(g_B g_C g_D)^3$ | $\pi$ | 48 |
| $A$ | $g_B(g_C g_D)^2$ | $\pi/2$ | 8 |
| $B$ | $g_C g_D g_A$ | $\pi/3$ | 6 |
| $B$ | $(g_C g_D g_A)^3$ | $\pi$ | 48 |
| $B$ | $(g_C g_D)^2 g_A$ | $\pi/2$ | 8 |
| $C$ | $g_D g_A g_B$ | $\pi/5$ | 10 |
| $C$ | $(g_D g_A g_B)^3$ | $3\pi/5$ | 10 |
| $C$ | $(g_D g_A g_B)^5$ | $\pi$ | 120 |
| $C$ | $g_D(g_A g_B)^2$ | $\pi/3$ | 6 |
| $D$ | $g_A g_B g_C$ | $\pi/5$ | 10 |
| $D$ | $(g_A g_B g_C)^3$ | $3\pi/5$ | 10 |
| $D$ | $(g_A g_B g_C)^5$ | $\pi$ | 120 |
| $D$ | $(g_A g_B)^2 g_C$ | $\pi/3$ | 6 |

Table 2: Inverse elliptic conjugacy classes

We will now show that *in the case $n = 4$ there are exactly three $\Gamma$-conjugacy classes of inverse elliptic elements, and in the case $n = 5$ there are exactly four classes,* compare table 2.

Because of symmetry reasons we only have to look at the corners $A$ and $C$. Firstly it is easy to see that there is a one-to-one correspondence between $\Gamma$-conjugacy classes of inverse elliptic elements of $\Gamma$ belonging to corner point $A$ and $\mathcal{E}_4$-conjugacy classes of inverse elliptic elements of $\mathcal{E}_4$. In the same way we have a one-to-one correspondence between $\Gamma$-conjugacy classes of inverse elliptic elements of $\Gamma$ belonging to corner point $C$ and $\mathcal{E}_5$-conjugacy classes of inverse elliptic elements of $\mathcal{E}_5$. (This is not necessarily true, e.g., for classes of plane reflections.)

Consider $n = 4$. To be sure that we have not missed an inverse elliptic conjugacy class we just count the elements in each class to see if the total number yields 24 (=number of orientation reversing elements in $\mathcal{E}_4$) minus the the number of elements in conjugacy classes of plane reflections. There are exactly two $\mathcal{E}_4$-conjugacy classes $\{g_i\}_{\mathcal{E}_4}$ and $\{g_j\}_{\mathcal{E}_4}$ of plane reflections, containing 3 and 6 elements, respectively: Remember that the number of elements in a conjugacy class $\{g\}_{\mathcal{E}_4}$ equals the index ord$\mathcal{E}_4$/ord$\mathcal{C}_{\mathcal{E}_4}(g)$ of the centralizer $\mathcal{C}_{\mathcal{E}_4}(g)$ in $\mathcal{E}_4$, which is $48/16 = 3$ (resp. $48/8 = 6$) in our case. 16 (resp. 8) is the order of the centralizer and corresponds to the angle $\pi/4$ (resp. $\pi/2$) at corner $A_2$ (resp. $A_1$) of our hexagon, as there are exactly 16 (resp. 8) elements of $\mathcal{E}_4$ that commute with $g_i$ (resp. $g_j$), namely the identity, 3 (resp. 1) rotations with axis perpendicular to the reflection plane, 4 (resp. 2) half turns with axis in the reflection plane, and all products of those elements with the considered reflection.

The elements $g_i g_j g_k$, $(g_i g_j g_k)^3$ and $g_j(g_k g_i)^2$ are not conjugate to one another, the orders of their centralizers are 6, 48 and 8. Hence their conjugacy classes contain 8, 1, 6 elements, respectively. Now $(3+6)+(8+1+6) = 24$, hence we have not missed any conjugacy class.

In the case $n = 5$ we have one conjugacy class of plane reflections, which contains $120/8 = 15$ elements. The centralizers of $g_i g_j g_k$, $(g_i g_j g_k)^3$, $(g_i g_j g_k)^5$ and $g_j(g_k g_i)^2$ have orders 10, 10, 120 and 6. $(15) + (12 + 12 + 1 + 20) = 60$, which is exactly the number of orientation reversing elements in $\mathcal{E}_5$. This completes the proof.

To give a first impression of the use of our trace formula (25) we will derive an average law for the *spectral staircase*

$$\mathcal{N}(E) = N + \sum_{n=N+1}^{\infty} \theta(p - p_n), \qquad p = \sqrt{E-1}, \qquad E > 1, \tag{33}$$

using only the geometric data discussed so far. ($N$ denotes the number of "small" eigenvalues, see section II.) We introduce the *smeared spectral staircase*

$$\mathcal{N}^{\epsilon}(E) = \sum_{n=1}^{\infty} h_{\epsilon}(p_n), \tag{34}$$

with the function

$$h_{\epsilon}(q) = \frac{1}{2}\left[\mathrm{erf}\left(\frac{p-q}{\epsilon}\right) + \mathrm{erf}\left(\frac{p+q}{\epsilon}\right)\right], \qquad \epsilon > 0, \tag{35}$$

satisfying conditions i-iii of section II. We split the trace formula expression of $\mathcal{N}^{\epsilon}(E)$ into its average part

$$\begin{aligned}\overline{\mathcal{N}}^{\epsilon}(E) :=& \frac{\mathrm{Vol}(\mathcal{F}_{\Gamma})}{6\pi^2}\, p^3 \\ &+ \frac{1}{8\pi} \sum_{\substack{\{\rho\}_{\Gamma} \\ \text{inv.}}} \chi(\rho)\, \mathrm{Area}(\mathcal{P}_{\rho}) \int_0^{\infty} dq\; q\, h_{\epsilon}(q)\, \coth(\pi q) \\ &+ \frac{1}{2\pi} \sum_{\substack{\{\rho\} \\ \text{ellipt.}}} \frac{\chi(\rho)\, l_{\tau_0}}{\mathrm{ord}\mathcal{E}(\rho)\,(1-\cos\phi_{\rho})}\, p \\ &+ \frac{1}{2} \sum_{\substack{\{\rho\}_{\Gamma} \\ \text{inv. ellipt.} \\ 0<\phi<\pi}} \frac{\chi(\rho)}{\mathrm{ord}\mathcal{E}_{\Gamma}(\rho)} \int_0^{\infty} dq\; h_{\epsilon}(q)\, \frac{\sinh[(\pi-\phi_{\rho})q]}{\sinh(\pi q)\, \sin\phi_{\rho}} \\ &+ \frac{1}{2} \sum_{\substack{\{\rho\}_{\Gamma} \\ \text{inv. ellipt.} \\ \phi=\pi}} \frac{\chi(\rho)}{\mathrm{ord}\mathcal{E}_{\Gamma}(\rho)} \int_0^{\infty} dq\; h_{\epsilon}(q)\, \frac{q}{\sinh(\pi q)}\end{aligned} \tag{36}$$

and its fluctuating part

$$\begin{aligned}\mathcal{N}^{\epsilon}_{\mathrm{fl}}(E) :=& \frac{1}{2\pi} \sum_{\substack{\{\tau\}_{\Gamma} \\ \text{hyperbol.}}} \frac{\chi(\tau)\, l_{\tau_0}}{\mathrm{ord}\mathcal{E}(\tau)\,(\cosh l_{\tau} - \cos\phi_{\tau})}\, \frac{\sin p l_{\tau}}{l_{\tau}}\, \exp\left(-\frac{\epsilon^2}{4} l_{\tau}^2\right) \\ &+ \frac{1}{2\pi} \sum_{\substack{\{\tau\}_{\Gamma} \\ \text{inv. hyperbol.}}} \frac{\chi(\tau)\, l_{\tau_0}}{\mathrm{ord}\mathcal{E}(\tau)\, \sinh l_{\tau}}\, \frac{\sin p l_{\tau}}{l_{\tau}}\, \exp\left(-\frac{\epsilon^2}{4} l_{\tau}^2\right).\end{aligned} \tag{37}$$

Performing the limit $\epsilon \to 0^+$ for $\overline{\mathcal{N}}^{\epsilon}(E)$ and neglecting exponentially small corrections we obtain a Weyl series in $p$:

$$\overline{\mathcal{N}}(E) = c_3 p^3 + c_2 p^2 + c_1 p + c_0 + O(\mathrm{e}^{-\pi p/5}), \qquad p \to \infty, \tag{38}$$

where

$$c_3 = \frac{\mathrm{Vol}(\mathcal{F}_\Gamma)}{6\pi^2}, \qquad c_2 = \frac{1}{16\pi} \sum_{\substack{\{\rho\}_\Gamma \\ \text{inv.}}} \chi(\rho)\, \mathrm{Area}(\mathcal{P}_\rho), \qquad c_1 = \frac{1}{2\pi} \sum_{\substack{\{\rho\}_\Gamma \\ \text{ellipt.}}} \frac{\chi(\rho)\, l_{\tau_0}}{\mathrm{ord}\mathcal{E}(\rho)\,(1-\cos\phi_\rho)},$$

$$c_0 = \frac{1}{96\pi} \sum_{\substack{\{\rho\}_\Gamma \\ \text{inv.}}} \chi(\rho)\, \mathrm{Area}(\mathcal{P}_\rho) + \frac{1}{4} \sum_{\substack{\{\rho\}_\Gamma \\ \text{inv. ellipt.}}} \frac{\chi(\rho)}{\mathrm{ord}\mathcal{E}_\Gamma(\rho)(1-\cos\phi_\rho)}.$$

This result holds for any co-compact lattice in $\mathfrak{H}_3$. Inserting the geometric data of $T_8$ and choosing Dirichlet boundary conditions (i.e., $\chi(g_i) = -1$ for all generators), we get $c_3 \approx 0.006056$, $c_2 = -3/32$, $c_1 = d(B,C)/8\pi + 2d(C,A)/9\pi + 5d(A,B)/32\pi + d(D,C)/5\pi \approx 0.456854$, $c_0 = -23/32$, see figure 7.

# IV The length spectrum of periodic orbits

The above derivation of the trace formula has shown the intimate relation between the classical periodic orbits and the conjugacy classes of the reflection group $\Gamma$ of the polyhedron. This suggests that the length spectrum of the classical periodic orbits can be computed by the conjugacy classes since a geometric method would be a too tantalizing effort in three dimensions. To this aim we use the generator product method, which has been used, e. g., in the computation of length spectra of hyperbolic octagons, which are surfaces being generated by Fuchsian groups [10]. This method generates all group elements $g \in \Gamma$ which can be represented by a product of the generators $g_A$, $g_B$, $g_C$ and $g_D$ having at most $N$ factors. Since the periodic orbits are directly connected with the conjugacy classes and not with the group elements, one has to count only group elements belonging to distinct conjugacy classes.

From definition (15) of conjugacy classes, it follows immediately that all cyclic permutations of a product belong to the same conjugacy class. Thus only the generator products are taken into account which are not cyclically equivalent to each other. If the reflection group $\Gamma$ was a free group this method would already yield all conjugacy classes. Unfortunately there are 10 identities between the generators by which the products can be transformed and which in turn have to be compared with respect to cyclic equivalence. For the considered tetrahedron these relations are

$$(39) \qquad g_A^2 = g_B^2 = g_C^2 = g_D^2 = \mathbf{1}$$

and

$$(40) \qquad (g_D\, g_A)^2 = (g_D\, g_B)^3 = (g_D\, g_C)^4 = (g_B\, g_C)^2 = (g_C\, g_A)^3 = (g_A\, g_B)^5 = \mathbf{1}$$

where $\mathbf{1}$ denotes the unit element. The computer program, which computes the length spectrum, uses symbolic algebra with four letters '1' up to '4' corresponding to the four generators $g_A, g_B, g_C, g_D$. With this choice of letters a map from the generator product onto integers is obtained by interpreting the word as a number. Then to every conjugacy class belongs a "minimal word" having the smallest number. The program starts with generating the words from the lowest possible number up to words with at most 20 letters. It then tries to reduce the word towards a word corresponding to a smaller number. If this is possible, the considered word belongs to a conjugacy class which has already been taken into account, otherwise the periodic orbit belonging to this word is stored. To reduce the word, all combinations of the identities (39) and (40) together with cyclic shifts have to be considered. Among the words with length 20 there are conjugacy classes with up to 2523 possible equivalent words. This contrasts to the earlier computation of the length spectrum of hyperbolic octagons where only one identity relation exists which only allows the reduction of very few words. Up to word length 20 there were 35 855 periodic orbits. It is worthwhile to note that this algorithm yields not only hyperbolic conjugacy classes but also elliptic and parabolic ones (if there are any). The only quantity in equation (14) being undetermined until now is $\mathrm{ord}\mathcal{E}_\Gamma(\tau)$. It is obtained by its geometric interpretation to be one for periodic orbits lying in the interior of the fundamental cell and to be two for surface orbits which do not lie along the edges of the fundamental cell. The order of edge orbits is given by the number of cells around the considered edge. Therefore, the computer program has to determine the geometric location of each orbit to get $\mathrm{ord}\mathcal{E}_\Gamma(\tau)$.

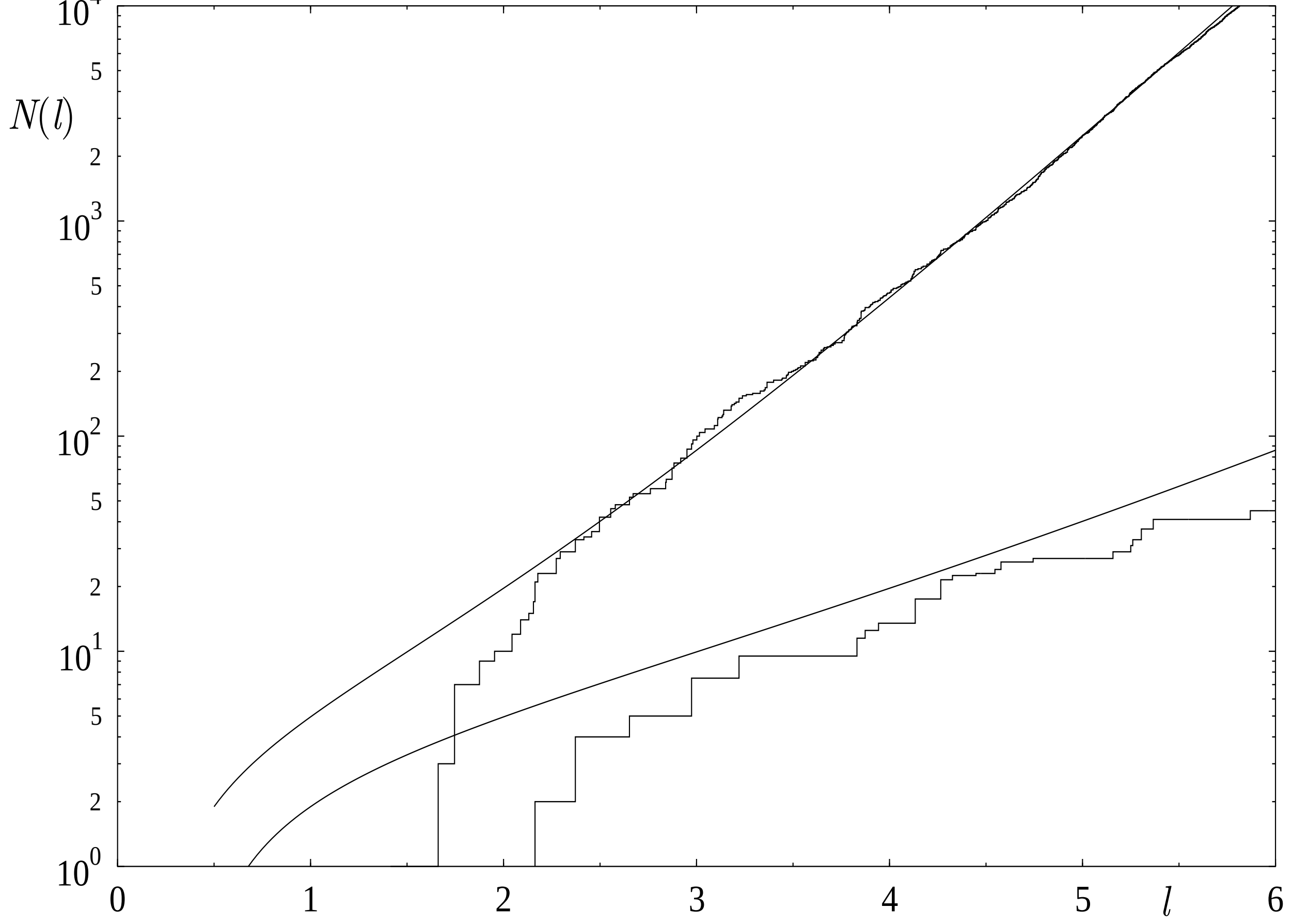

Figure 3: The classical staircase $N(l)$ is shown in comparison with the asymptotic behaviour (41) for $l < 6$. The lower staircase counts only surface orbits. It is displayed together with $\mathrm{Ei}(l)$.

The classical staircase $N(l)$ counts the number of primitive periodic orbits with length less than or equal to $l$. The asymptotic behaviour of $N(l)$ is given by [11]

$$(41) \qquad N(l) \;\sim\; \mathrm{Ei}(\tau l) \;\sim\; \frac{\exp(\tau l)}{\tau l}, \qquad l \to \infty,$$

where $\tau$ denotes the topological entropy, which in the case of finite hyperbolic spaces of constant curvature $K = -1$ is connected with the dimension $D$ via $\tau = D - 1$.

Since the topological entropy $\tau$ together with the mean level density $\overline{d}(E)$ determines the efficiency of periodic-orbit theory [12], a larger topological entropy implies that numerical appli-cations of periodic-orbit theory are much more difficult in three dimensions. In figure 3 (upper curve) the classical staircase $N(l)$ is shown in comparison with the asymptotic behaviour (41) with $\tau = 2$ for $l < 6$. This demonstrates that the computer program has provided the length spectrum up to $l = 6$ nearly complete. Indeed, the first deviation from the asymptotic behaviour occurs at $l \simeq 5.5$, which is the cut-off length used in the evaluation of the trace formula in the next section. In addition the staircase counting only surface orbits is also shown. Since the surface can be considered as a two-dimensional billiard, the topological entropy of the surface orbits is $\tau = 1$. The corresponding staircase is compared with $\mathrm{Ei}(l)$. Since surface orbits are represented by unusually long words in the code used, these orbits are not so completely computed as non-surface orbits. Thus their staircase lies somewhat below the curve belonging to $\mathrm{Ei}(l)$. This figure gives an impression of the wealth of periodic orbits in three dimensions in comparison with two dimensions.

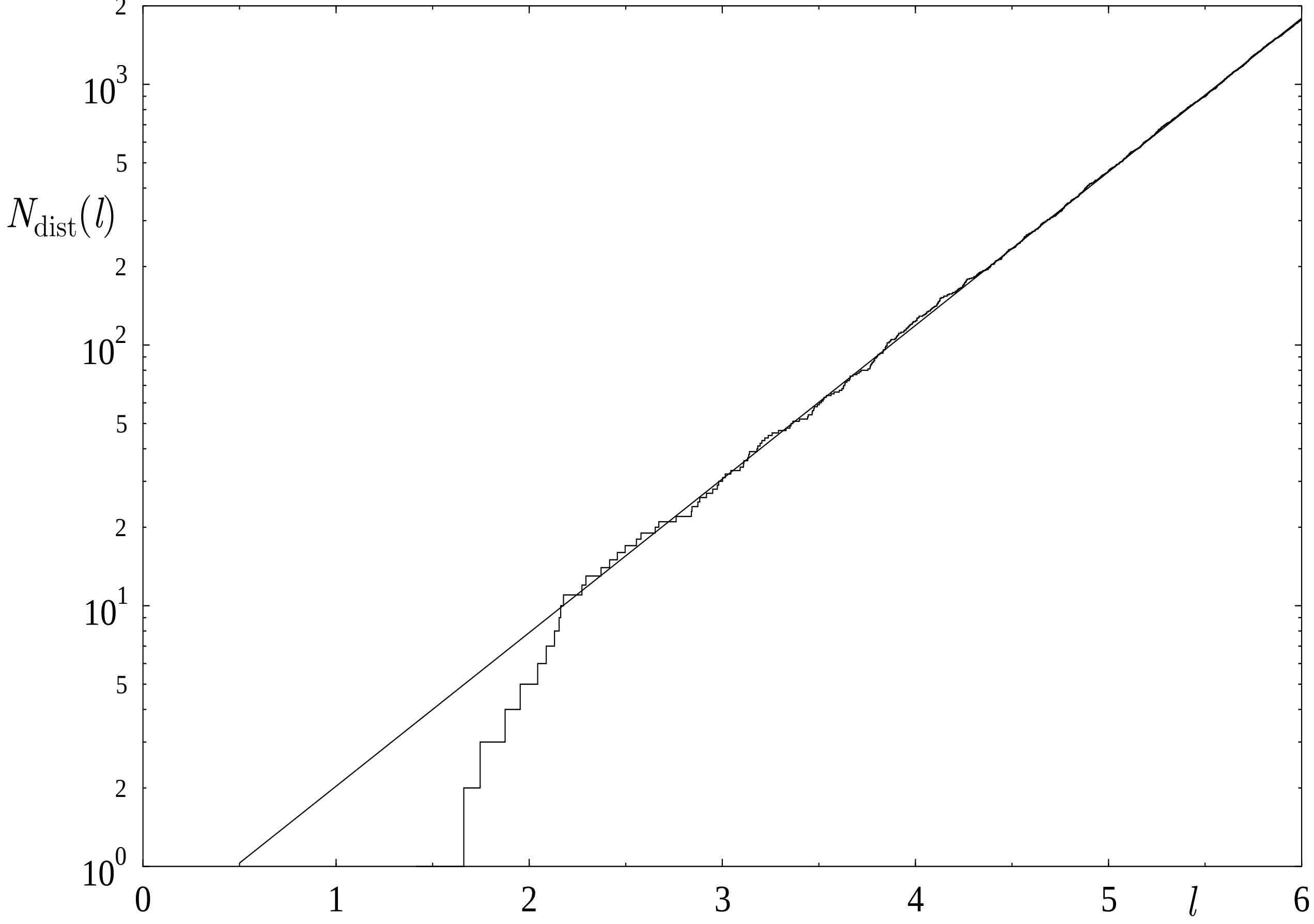

Figure 4: The staircase counting only the periodic orbits having distinct lengths is shown in comparison with the fit $N^{\text{f}}_{\text{dist}}(l) = a\mathrm{e}^{bl}$ with $a = 0.52286$ and $b = 1.35668$.

Since the tetrahedron $T_8$ is not arithmetic, one expects a generic length spectrum of periodic orbits, where the multiplicities of the lengths are determined only by the symmetries of the classical system. As the equations of motion are symmetric under time reversal, the multiplicity of a length should be one, two or four depending on whether the periodic orbit possesses a back traversal or not, and whether it is invariant with respect to the symmetry $\rho_\pi$. It is therefore very surprising to observe a length spectrum with exponentially increasing multiplicities as a function of length. This immediately reminds us of the eight relatives of the tetrahedron $T_8$, which are all arithmetic and thus possess an exponentially degenerated length spectrum with

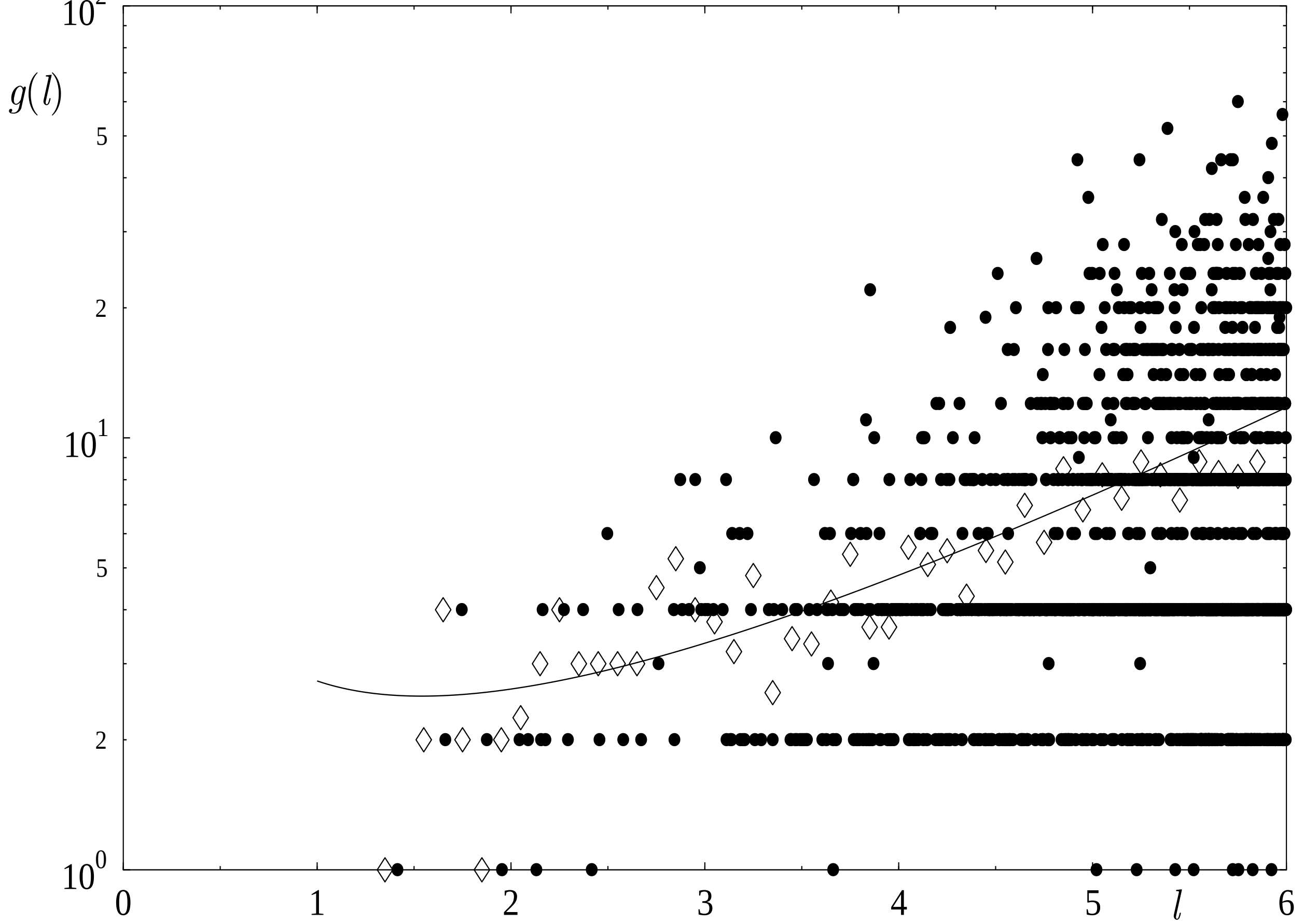

Figure 5: The strongly fluctuating multiplicities are displayed as dots. A local average of the multiplicities over $\Delta l = 0.1$ (diamonds) suppresses the fluctuations and is consistent with the mean behaviour following from the fit of $N_{\text{dist}}(l)$ (full curve).

a mean multiplicity

$$\langle g(l) \rangle \sim c_\Gamma \frac{\exp(l)}{l}, \qquad l \to \infty, \tag{42}$$

being a consequence of the slow increase of the number of distinct lengths

$$N_{\text{dist}}(l) \sim c_\Gamma^{-1} \exp(l). \tag{43}$$

For quaternion groups, whose trace field is invariant under complex conjugation, the constant is given by [13]

$$c_\Gamma = \frac{|D_{\mathfrak{a}}|^{1/2}}{2^{d-3}\pi}, \tag{44}$$

where $d$ is the degree of the trace field $K$ of $\Gamma$ over the rational numbers $\mathbb{Q}$, and $D_{\mathfrak{a}}$ the discriminant of the minimal ideal $\mathfrak{a}$ of $K$ that contains all traces of $\Gamma$.

Indeed the trace fields of all arithmetic reflection groups are invariant under complex conjugation, therefore they are commensurable to quaternion groups of the type described above, and should satisfy (42), with a different constant, however.

The exponential behaviour of multiplicities of arithmetic manifolds was detected in the two-dimensional case first for the regular octagon [10], and later predicted for more general

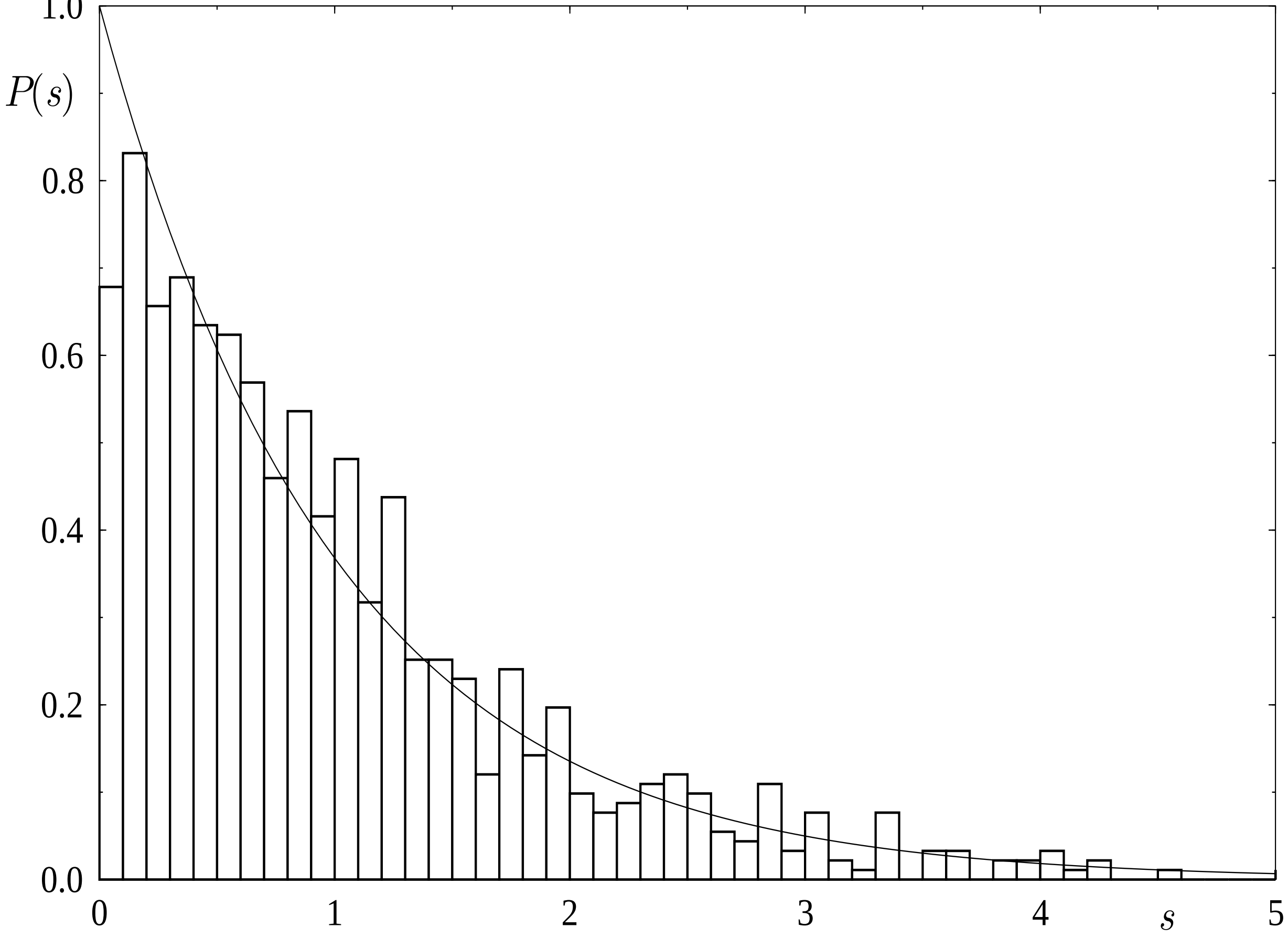

Figure 6: The length spacing distribution $P(s)$ of the unfolded length spectrum is shown in comparison with the Poisson distribution $P_{\mathrm{Poisson}}(s) = e^{-s}$. The histogram contains all spacings up to length $l = 5.5$.

arithmetic surfaces [14, 15, 16], where

$$\langle g(l) \rangle \;\sim\; c_\Gamma \, \frac{\exp(l/2)}{l}, \qquad l \to \infty, \tag{45}$$

with a constant explicitly known for quaternion groups and some special cases. For details see [16, 17]. We would like to note that the relation between the mean multiplicities for commensurable groups $\Gamma_a$, $\Gamma_b$ given in [16, 17] is not valid in general.

The spectral statistics of a chaotic system depend very sensibly on the behaviour of the multiplicities in the length spectrum. Quantum energy spectra of arithmetic surfaces even show a Poisson-like behaviour, which is typical for integrable systems and absolutely unexpected for chaotic systems [14, 15, 17]. Poisson-like statistics are also predicted for three-dimensional arithmetic systems [13]. Besides that, there are no other chaotic systems, for which there exist more rigorously proven results concerning quantum spectral properties, see [18, 19] for reference. We have decided to look at the more generic tetrahedron $T_8$, which is – as stated above – the only non-arithmetic one.

Since the expectation for a generic system is $N_{\mathrm{dist}}(l) \sim a\frac{\mathrm{e}^{2l}}{2l}$ , it is interesting to determine the kind of exponential growth. In figure 4 $N_{\mathrm{dist}}(l)$ is shown in comparison with a fit $N^{\mathrm{f}}_{\mathrm{dist}}(l) = a\mathrm{e}^{bl}$ with $a = 0.52286$ and $b = 1.35668$. A fit function with the asymptotic behaviour of $\mathrm{Ei}(al)$ could be excluded. Since a length spectrum of an arithmetic system would have yielded $b = 1$, the multiplicity is smaller for the tetrahedron $T_8$ but nevertheless exponential. The mean multiplicity corresponding to the fit is given by $\langle g_f(l) \rangle = \frac{\mathrm{e}^{(2-b)l}}{abl}$. The true multiplicities are strongly fluctuating as shown in figure 5. For that reason figure 5 also displays a local average over a length interval of $\Delta l = 0.1$ which reveals the increase of the multiplicities, and also the mean behaviour $\langle g_f(l) \rangle$. The strong fluctuations are reminiscent of an arithmetic system like the regular octagon for which the fluctuations of the multiplicities are shown in [10]. There remains to answer the crucial question, why the non-arithmetic tetrahedron $T_8$ displays properties similar to arithmetic systems.

At the end of this section we want to discuss the length spacing distribution $P(s)$. This quantity was originally studied in connection with the quantal level spectrum and is known as nearest neighbour level spacing distribution. The length spacing distribution was first studied in [20] for the hyperbola billiard which possesses a length spectrum being generic in the sense that the multiplicities are in accordance with the symmetries. In [20] it was shown that $P(s)$ displays a Poisson distribution. In figure 6 the length spacing is shown for the tetrahedron $T_8$ where all length spacings up to length $l = 5.5$ are included. The distribution roughly agrees with a Poisson distribution, and one observes a length clustering since the distribution has its maximum near $s = 0$. Thus in this three-dimensional billiard the length spacing agrees empirically with the one of the hyperbola billiard despite its anomalous behaviour with respect to the multiplicities.

# V The spectral staircase as a quantization tool

In this section we would like to demonstrate as an application of the trace formula (25) the computation of the spectral staircase $\mathcal{N}(E)$ again for the tetrahedron $T_8$ with Dirichlet boundary conditions. The main motivation for computing the spectral staircase $\mathcal{N}(E)$ arises from the fact that it provides a very useful quantization rule.

In order to compare the results obtained from classical quantities with the quantum mechanical ones, the quantal level spectrum has to be computed. To this aim the boundary element method is employed. Define the following differential operator $\mathcal{L}_x := -\Delta_x - E$, $x \in \mathbb{R}^3$, then the Schrödinger equation (4) reads

$$\mathcal{L}_x \psi(x) = 0 \quad . \tag{46}$$

The free Green function given in equation (10) satisfies

$$\mathcal{L}_x \, G(x, y; E) = \delta(x - y) \quad , \tag{47}$$

with $\delta(x)$ being the three-dimensional Dirac delta distribution. Let $\Sigma$ be the surface of the polyhedron, then a solution $\psi(x)$ of the Schrödinger equation can be represented as a surface integral for $x \notin \Sigma$

$$\psi(x) = \int_\Sigma d\Sigma_y \, G(x, y; E) \, \Phi(y) \quad , \tag{48}$$

as is verified by applying $\mathcal{L}_x$ on equation (48). The function $\Phi(y)$ with $y \in \Sigma$ has to be determined, such that $\psi(x)$ satisfies the boundary conditions for $x \in \Sigma$. Expanding $\Phi$ in terms of simple ansatz functions, a set of linear equations is obtained from (48) in terms of the expansion coefficients $c_i$ of $\Phi$. The Dirichlet boundary conditions are imposed by setting $\psi(x) = 0$ for $x \in \Sigma$. The resulting system of equations $M_E \vec{c} = 0$ has non-trivial solutions only if the determinant of the energy dependent matrix $M_E$ vanishes, i.e., if $E$ corresponds to a quantal level $E_n$. Since the matrix $M_E$ is nearly singular, a singular value decomposition is applied to $M_E$. Then the zeros of the singular values betray the locations of the quantal levels. To obtain the quantal levels with respect to the two symmetry classes of the tetrahedron $T_8$ the Green function $G^{\pm}(x, y; E) = G(x, y; E) \pm G(x, \rho_\pi(y); E)$ is used. All quantal levels in the energy interval $E \in [0, 500]$ have been computed. For both classes 31 quantal levels are found.

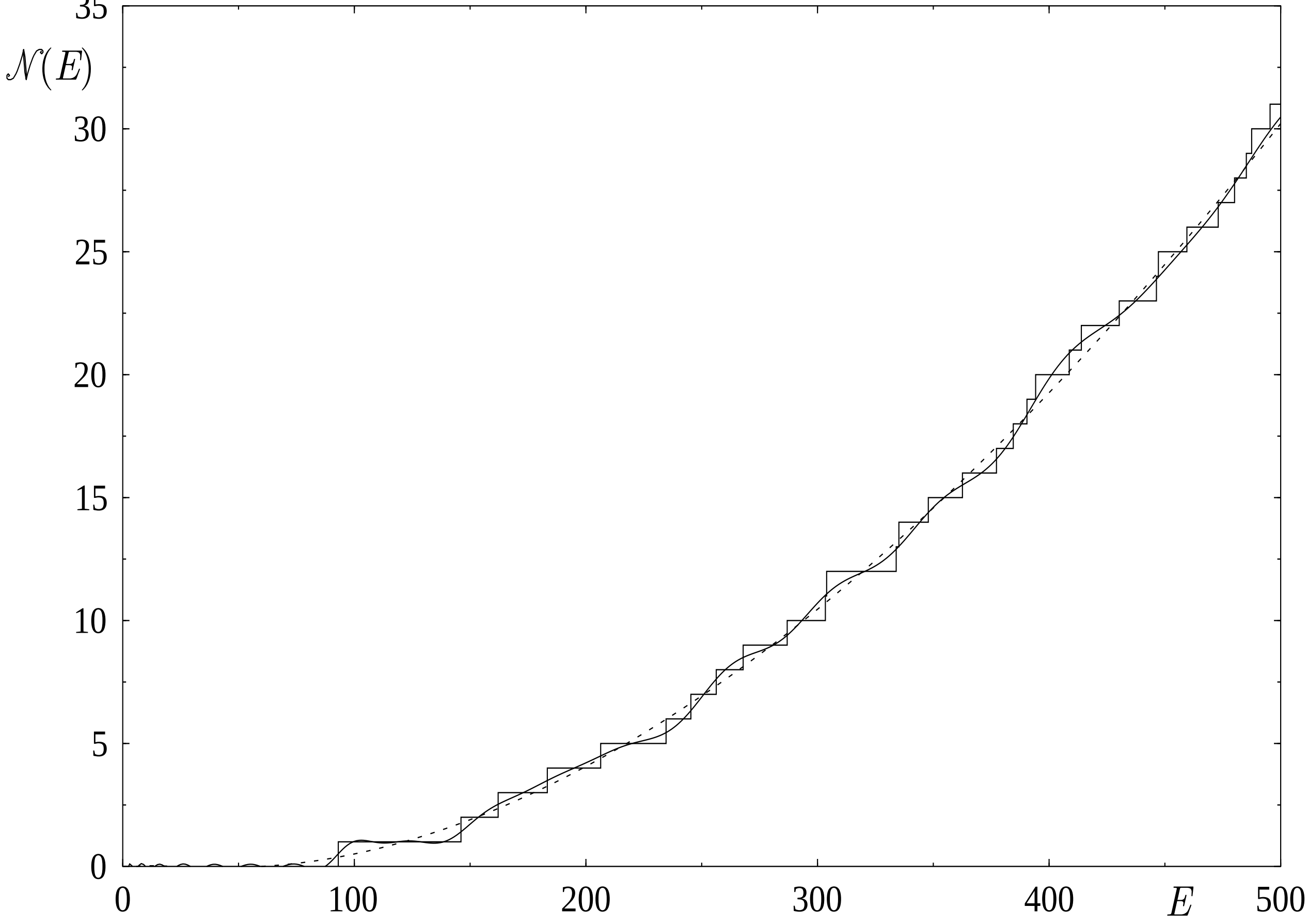

Figure 7: The spectral staircase $\mathcal{N}(E)$ is shown in comparison with the average part $\overline{\mathcal{N}}(E)$ (dotted curve) and the full periodic-orbit theory (full curve), where in the latter all periodic orbits with length $l < 5.5$ are taken into account.

Let us now turn to the computation of $\mathcal{N}(E)$ in terms of the classical quantities. The asymptotic behaviour $E \to \infty$ is already determined without the hyperbolic conjugacy classes and has been derived in section III, see equation (36). The fluctuating part $\mathcal{N}_{\text{fl}}(E)$, defined in equation (37), due to hyperbolic conjugacy classes, i. e., due to the periodic orbits, describes the deviations of $\mathcal{N}(E)$ from the average part (37). The series over the hyperbolic conjugacy classes in (37) are absolutely convergent for $\epsilon > 0$. However, to extract from the periodic orbits as much information as possible, the smallest possible smoothing parameter $\epsilon$ should be used. For $\epsilon = 0$ the series are not absolutely convergent, however, they may be conditionally convergent. The numerical evaluation of the periodic-orbit sums up $l = 5.5$ indicates that the characters $\chi$ attached to the periodic orbits are distributed in a sufficiently random way such that both sums are conditionally convergent. In figure 7 the spectral staircase $\mathcal{N}(E)$ is evaluated for $\epsilon = 0$ using the length spectrum up to $l = 5.5$ (full curve). In addition, the average part $\overline{\mathcal{N}}(E)$ is shown (dotted curve) also for $\epsilon = 0$ which differs from the Weyl series (38) only by exponentially small terms. Both curves can be compared with the "true" spectral staircase using the quantal levels obtained by the boundary element method. Surprisingly, the average part $\overline{\mathcal{N}}(E)$ describes the mean behaviour of $\mathcal{N}(E)$ down to the lowest state, an observation made also in many other billiard systems. The contributions of the periodic orbits with length $l < 5.5$ already conspire

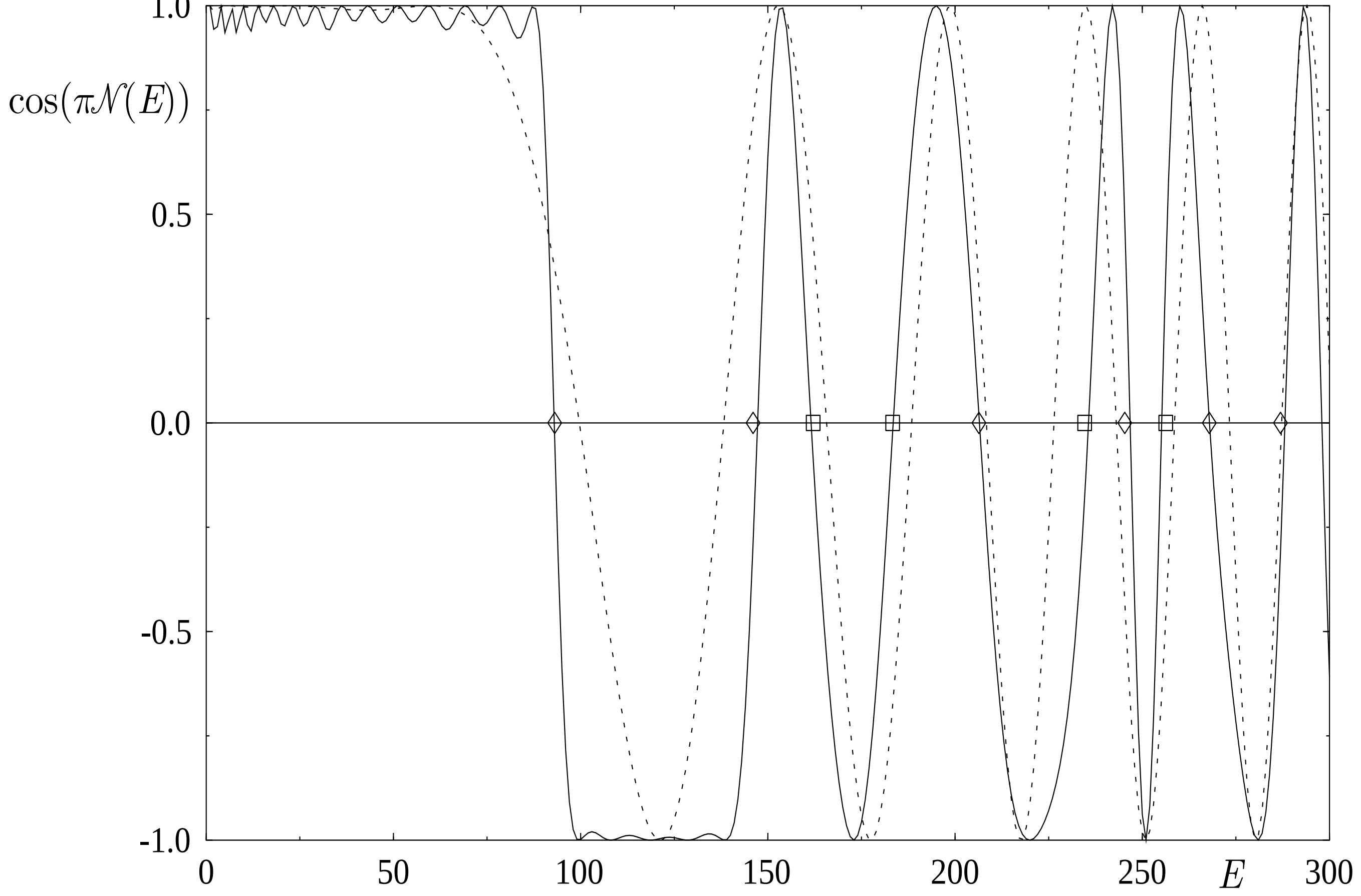

Figure 8: The quantization function $\cos(\pi\mathcal{N}(E))$ (full curve) is shown together with the quantal levels shown as diamonds and squares for positive and negative symmetry class, respectively. The dotted curve belongs to $\cos(\pi\overline{\mathcal{N}}(E))$

to yield the first step where the curve lies around one in the range $E \simeq 100$ to $E \simeq 140$. Steps corresponding to higher excited states are not reproduced. The contributions of periodic orbits with lengths $l > 5.5$ are required to generate the higher steps. However, a close look at figure 7 reveals that for not too highly excited levels the full curve crosses the steps at almost half height, which is not the case for the average part $\overline{\mathcal{N}}(E)$. This behaviour is important for the application of the spectral staircase as a quantization tool.

In [21, 22] an approximation of the spectral staircase using only geometric quantities has been employed to quantize three different billiard systems by using the quantization condition

$$\cos(\pi\,\mathcal{N}(E)\,) \;=\; 0 \quad . \tag{49}$$

The name quantization condition is justified since only classical quantities enter the equations (36) and (37). In figure 8 the function $\cos(\pi\mathcal{N}(E)\,)$ is shown using the approximated staircase computed with the length spectrum up to $l = 5.5$ (full curve), and it is seen that its zeros agree very well with the first quantal levels indicated by diamonds and squares for the positive and negative $\rho_\pi$-symmetry class, respectively. It may be surprising that the levels are resolved so well by keeping in mind that the staircase itself is only generated for the first level, but, as mentioned above, this quantization condition already works fine if the steps are crossed at half height. The quantization obtained using only the average part $\overline{\mathcal{N}}(E)$ is shown in figure 8 as a dotted curve. One observes, that all zeros correspond to a level being a consequence of using the average part $\overline{\mathcal{N}}(E)$. Furthermore, the improvement due to the contributions of periodic orbits with length up to $l = 5.5$ is impressive.

# VI Summary

This paper is devoted to the study of strongly chaotic compact systems embedded in the three-dimensional hyperbolic space having constant negative curvature. Special emphasis lies on the relation between the classical and the quantum mechanical aspects being exemplified by a trace formula which allows to express a sum over quantal levels in terms of purely geometric quantities. In the case of two-dimensional systems it was very valuable to have the exact Selberg trace formula for various comparisons with Gutzwiller's semiclassical periodic-orbit theory [2]. It is thus highly desirable to have exact trace formulae for analogous comparisons in three dimensions. The first step in this direction has been undertaken in [23] where the trace of the regularized resolvent has been derived for compact systems which tessellate the three-dimensional hyperbolic space corresponding to billiard systems with periodic boundary conditions. The derivation of the trace formula is based on the lattice group which realizes the tessellation of the hyperbolic space in terms of the fundamental cell. In [23] the group contains only orientation preserving elements. The trace of the resolvent for the more general case including orientation reversing elements is derived in this paper. The trace of the resolvent (14) is generalized to the trace formula (25) applicable to a vast class of test functions. The inclusion of orientation reversing elements has the advantage that in addition to the periodic boundary conditions also Dirichlet and Neumann boundary conditions are covered by this trace formula. An exact trace formula has thus been found for all compact billiard systems which tessellate the hyperbolic space.

There are three-dimensional models which have been proposed in [1] for a study of quantum chaos. However, they are too complicated for a first application of the trace formula. The simplest three-dimensional systems are provided by tetrahedra out of which nine are possible in hyperbolic space. Eight of these are arithmetic and thus lead to non-generic quantum systems. Therefore, we have chosen the only non-arithmetic tetrahedron, called $T_8$. The conjugacy classes of $T_8$ have been discussed in detail and the length spectrum of periodic orbits, connected with hyperbolic and inverse hyperbolic conjugacy classes, has been numerically computed. The numerical result indicates that the multiplicities of periodic orbits with identical lengths increase exponentially with length, a result unexpected for a non-arithmetic system. The unusual properties of arithmetic quantum chaotic systems are ascribe to these exponentially increasing multiplicities. The computed length spectrum allows an approximate computation of the spectral staircase $\mathcal{N}(E)$ which in turn provides a valuable quantization tool, see equation (49). In order to compare these results directly with quantum mechanics, the quantal levels up to $E = 500$ for $T_8$ with Dirichlet boundary conditions have been computed using the boundary element method. The first quantization of a three-dimensional chaotic system using periodic-orbit theory is displayed in figure 8 where the excellent agreement of the zeros of $\cos(\pi\mathcal{N}(E))$ with the quantal levels is seen. This is especially impressive because the topological entropy of three-dimensional systems in hyperbolic space is $\tau = 2$ being the highest topological entropy ever encountered in numerical applications of trace formulae of chaotic systems.

# Appendix: Proof of the trace formula

Elstrodt, Grunewald and Mennicke [23] have derived a trace formula for co-compact Kleinian groups, i.e., co-compact lattices of $\mathrm{Iso}^+\mathfrak{H}_3 \simeq \mathrm{PSL}(2,\mathbb{C})$. To extend their result to general co-compact lattices $\Gamma$ of $\mathrm{Iso}\,\mathfrak{H}_3$, suppose $\Gamma$ is not a Kleinian group. It can be shown that every such lattice $\Gamma$ contains a Kleinian subgroup of index two, namely the group $\Gamma^+$ of all orientation preserving elements of $\Gamma$. We have the coset decomposition

$$\Gamma = \Gamma^+ \cup \Gamma^+\sigma. \tag{50}$$

A fundamental cell $\mathcal{F}_{\Gamma^+}$ can be chosen in such a way that it consists of two copies of the fundamental cell of $\Gamma$:

$$\mathcal{F}_{\Gamma^+} = \mathcal{F}_\Gamma \cup \sigma(\mathcal{F}_\Gamma). \tag{51}$$

As the integration in (12) is independent of a particular choice of the fundamental cell, we could as well integrate over $\sigma(\mathcal{F}_\Gamma)$, or equivalently over the double cell $\mathcal{F}_{\Gamma^+}$ such that

$$\int_{\mathcal{F}_\Gamma} d\mu(x)\ [G_\Gamma(x,x;E) - G_\Gamma(x,x;E')] = \frac{1}{2}\int_{\mathcal{F}_{\Gamma^+}} d\mu(x)\ [G_\Gamma(x,x;E) - G_\Gamma(x,x;E')]. \tag{52}$$

For the rest of this derivation suppose $\operatorname{Im} p > 1$, $\operatorname{Im} p' > 1$. We write $G_\Gamma(x,x;E)$ as the coherent superposition (9). We can carry out the same estimate as in [23] to see that we are permitted to exchange sum and integration. We then obtain

$$\begin{aligned} &\frac{1}{2}\sum_{g\in\Gamma}\ \int_{\mathcal{F}_{\Gamma^+}} d\mu(x)\ \chi(g)\ [G(g(x),x;E) - G(g(x),x;E')] \\ &\qquad = \frac{1}{2}\sum_{g\in\Gamma^+}\ \int_{\mathcal{F}_{\Gamma^+}} d\mu(x)\ \chi(g)\ [G(g(x),x;E) - G(g(x),x;E')] \\ &\qquad\qquad + \frac{1}{2}\sum_{g\in\Gamma^+\sigma}\ \int_{\mathcal{F}_{\Gamma^+}} d\mu(x)\ \chi(g)\ [G(g(x),x;E) - G(g(x),x;E')]. \end{aligned} \tag{53}$$

The sum over orientation preserving elements of $\Gamma$ has been worked out in [23]; the result reads

$$\begin{aligned} &\frac{1}{2}\sum_{g\in\Gamma^+}\ \int_{\mathcal{F}_{\Gamma^+}} d\mu(x)\ \chi(g)\ [G(g(x),x;E) - G(g(x),x;E')] \\ &\qquad = -\frac{\operatorname{Vol}(\mathcal{F}_{\Gamma^+})}{8\pi \mathrm{i}}(p-p') \\ &\qquad\quad - \sum_{\substack{\{\rho\}_{\Gamma^+} \\ \text{ellipt.}}} \frac{\chi(\rho)\, l_{\tau_0^+}}{8\,\mathrm{ord}\mathcal{E}_{\Gamma^+}(\rho)\,(1-\cos\phi_\rho)} \left(\frac{1}{\mathrm{i}p} - \frac{1}{\mathrm{i}p'}\right) \\ &\qquad\quad - \sum_{\substack{\{\tau\}_{\Gamma^+} \\ \text{hyperbol.}}} \frac{\chi(\tau)\, l_{\tau_0^+}}{8\,\mathrm{ord}\mathcal{E}_{\Gamma^+}(\tau)\,(\cosh l_\tau - \cos\phi_\tau)} \left(\frac{\exp(\mathrm{i}\,p\,l_\tau)}{\mathrm{i}p} - \frac{\exp(\mathrm{i}\,p'\,l_\tau)}{\mathrm{i}p'}\right). \end{aligned} \tag{54}$$

$\tau_0^+$ is a hyperbolic element of $\Gamma^+$ with minimal length $l_{\tau_0^+}$ that commutes with $\rho$ or $\tau$, respectively. For further explanations see section II, just replace $\Gamma$ by $\Gamma^+$.

Let us turn to the second sum over orientation reversing elements. Using Selberg's trick [24] we rewrite (53) as a sum over $\Gamma^+$-conjugacy classes $\{g\}_{\Gamma^+}$ of orientation reversing elements $g \in \Gamma - \Gamma^+$, where the integration has to be performed over the fundamental cell of the $\Gamma^+$-centralizer $\mathcal{C}_{\Gamma^+}(g)$,

$$\frac{1}{2}\sum_{\substack{\{g\}_{\Gamma^+} \\ g\in\Gamma-\Gamma^+}}\ \int_{\mathcal{F}_{\mathcal{C}_{\Gamma^+}(g)}} d\mu(x)\ \chi(g)\ [G(g(x),x;E) - G(g(x),x;E')] \tag{55}$$

There are three kinds of conjugacy classes we have to deal with: classes of plane reflections, of inverse elliptic elements (we distinguish between point reflections and other inverse elliptic elements) and of inverse hyperbolic elements.

**Classes of plane reflections.** Each plane reflection $\rho$ is $\mathrm{PSL}(2,\mathbb{C})$-conjugate to the reflection $\mathbf{1}\mathrm{j}$ at the $(x_1,x_3)$-plane, i.e., there is a $g \in \mathrm{PSL}(2,\mathbb{C})$, such that $\mathrm{j} = g\rho g^{-1}$. We have

$$(56)\quad \frac{\chi(\rho)}{2} \int\limits_{\mathcal{F}_{\mathcal{C}_{\Gamma^+}(\rho)}} d\mu(x)\; [G(\rho(x),x;E) - G(\rho(x),x;E')]$$

$$= \frac{\chi(\rho)}{2} \int\limits_{\mathcal{F}_{\mathcal{C}_\Theta(\mathrm{j})}} d\mu(x)\; [G(\mathrm{j}(x),x;E) - G(\mathrm{j}(x),x;E')],$$

where $\Theta := g\Gamma^+ g^{-1}$. It follows from

$$\mathrm{j}\begin{pmatrix} a & b \\ c & d \end{pmatrix} = \pm \begin{pmatrix} a & b \\ c & d \end{pmatrix}\mathrm{j}$$

that all coefficients $a,b,c,d$ are either real or imaginary, hence the centralizer $\mathcal{C}_\Theta(\mathrm{j})$ is the subgroup of all elements contained both in $\Gamma^+$ and

$$\mathrm{PSL}(2,\mathbb{R}) \cup \mathrm{PSL}(2,\mathbb{R}) \begin{pmatrix} \mathrm{i} & 0 \\ 0 & -\mathrm{i} \end{pmatrix}.$$

$\mathcal{C}_\Theta(\mathrm{j})$ tessellates the $(x_1,x_3)$-plane, and can therefore be viewed as a two-dimensional lattice, which is naturally co-compact. By $\mathcal{P}_\rho$ we denote its fundamental cell on the $(x_1,x_3)$-plane (which in general is a hyperbolic polygon).

The fundamental cell $\mathcal{F}_{\mathcal{C}_\Theta(\mathrm{j})}$ of the centralizer $\mathcal{C}_\Theta(\mathrm{j})$ in $\mathfrak{H}_3$ has the shape

$$(57)\quad \mathcal{F}_{\mathcal{C}_\Theta(\mathrm{j})} = \{x \in \mathfrak{H}_3 \mid (r,z) \in \mathcal{P}_\rho,\ \varphi \in (-\pi/2, \pi/2)\},$$

with cylindric coordinates

$$x_1 = z, \qquad x_2 = r\sin\varphi, \qquad x_3 = r\cos\varphi.$$

Hence (56) equals

$$(58)\quad \frac{\chi(\rho)}{8\pi} \iint\limits_{\mathcal{P}_\rho} \frac{dz\,dr}{r^2} \int\limits_{-\pi/2}^{\pi/2} \frac{d\varphi}{\cos^3\varphi}\, \frac{\exp[\mathrm{i}\,p\,d(\mathrm{j}(x),x)] - \exp[\mathrm{i}\,p'\,d(\mathrm{j}(x),x)]}{\sinh d(\mathrm{j}(x),x)},$$

where

$$d(\mathrm{j}(x),x) = 2\,\mathrm{arsinh}\frac{|x_2|}{x_3} = 2\,\mathrm{arsinh}|\tan\varphi|.$$

The $(z,r)$-integration yields the area of $\mathcal{P}_\rho$, the $\varphi$-integration can be transformed into a representation of the psi function

$$\psi(x) := \frac{\Gamma'}{\Gamma}(x).$$

We finally get

$$(59)\quad -\chi(\rho)\,\frac{\mathrm{Area}(\mathcal{P}_\rho)}{16\pi}\,[\psi(1-\mathrm{i}p) + \psi(-\mathrm{i}p) - \psi(1-\mathrm{i}p') - \psi(-\mathrm{i}p')].$$

**Classes of point reflections.** Choosing suitable coordinates, the point reflection $\rho$ corresponds to the reflection $\tilde{\rho}$ at point $x = \mathrm{j}$, i.e.,

$$\tilde{\rho} := \begin{pmatrix} 0 & \mathrm{j} \\ -\mathrm{j} & 0 \end{pmatrix} = g\rho g^{-1},\ g \in \mathrm{PSL}(2,\mathbb{C}),$$

with suitable $g \in \mathrm{PSL}(2,\mathbb{C})$. Again let $\Theta := g\Gamma^+ g^{-1}$.

*The centralizer $\mathcal{C}_\Theta(\tilde{\rho})$ consists of all elements of finite order in $\Theta$ that leave the point $x = \mathrm{j}$ invariant.*

To proof this statement consider the equation

$$\begin{pmatrix} 0 & \mathrm{j} \\ -\mathrm{j} & 0 \end{pmatrix}\begin{pmatrix} a & b \\ c & d \end{pmatrix} = \pm\begin{pmatrix} a & b \\ c & d \end{pmatrix}\begin{pmatrix} 0 & \mathrm{j} \\ -\mathrm{j} & 0 \end{pmatrix},$$

hence the coefficients must fulfill the relations $a = \pm\overline{d}$ and $b = \mp\overline{c}$, therefore

$$\begin{pmatrix} a & b \\ c & d \end{pmatrix} = \pm\begin{pmatrix} a & b \\ \mp\overline{b} & \pm\overline{a} \end{pmatrix}.$$

The lower sign is ruled out as the determinant would be negative. The trace is real and contained in the interval $(-2,2)$, as $|a|^2+|b|^2 = 1$, therefore $|a| \leq 1$, hence $|\mathrm{Re}\,a| \leq 1$, equality if and only if $a = \pm 1$, $b = 0$.

Point j is invariant under such transformations, because

$$(a\mathrm{j}+b)(-\overline{b}\mathrm{j}+\overline{a})^{-1} = \mathrm{j}.$$

□

$\mathcal{C}_\Theta(\tilde{\rho})$ is finite (as $\Theta$ is discrete) and therefore trivially coincides with its maximal finite subgroup $\mathcal{E}_\Theta(\tilde{\rho})$. By conjugation the same is true for $\mathcal{C}_{\Gamma^+}(\rho) = \mathcal{E}_{\Gamma^+}(\rho)$.

Defining

$$I(p,\phi) := \frac{1}{4\pi}\int_{\mathfrak{H}_3} d\mu(x)\,\frac{\exp[\mathrm{i}\,p\,d(\rho(x),x)]}{\sinh d(\rho(x),x)},\ \mathrm{Im}\,p > 1, \tag{60}$$

where

$$\rho = \begin{pmatrix} 0 & \mathrm{i}e^{\mathrm{i}\phi/2} \\ \mathrm{i}e^{-\mathrm{i}\phi/2} & 0 \end{pmatrix}\mathrm{j},$$

we have

$$\frac{1}{2}\int_{\mathcal{F}_{\mathcal{C}_\Theta(\rho)}} d\mu(x)\ \chi(\rho)\,[G(\rho(x),x;E) - G(\rho(x),x;E')] = \frac{\chi(\rho)}{2\,\mathrm{ord}\mathcal{E}_\Theta(\rho)}\,[I(p,\pi) - I(p',\pi)]. \tag{61}$$

Evaluating the integral in definition (60) of $I(p,\phi)$ we get

$$I(p,\phi) = \begin{cases} \dfrac{1}{4\mathrm{i}p} + \dfrac{\beta(-\mathrm{i}p)}{2} & \text{if } \phi = \pi \mod 2\pi \\ \dfrac{1}{2\mathrm{i}p\sin\phi}\displaystyle\sum_{k=1}^{\infty}\frac{\sin k\phi}{\mathrm{i}p-k} & \text{if } \phi \neq 0 \mod \pi, \end{cases} \tag{62}$$

with

$$\beta(x) = \frac{1}{2}\left[\psi\left(\frac{x+1}{2}\right) - \psi\left(\frac{x}{2}\right)\right].$$

**Classes of further inverse elliptic elements.** $\rho$ is conjugate to

$$\tilde{\rho} := \begin{pmatrix} 0 & \mathrm{i}e^{\mathrm{i}\phi/2} \\ \mathrm{i}e^{-\mathrm{i}\phi/2} & 0 \end{pmatrix} \mathrm{j} = g\rho g^{-1}, \; g \in \mathrm{PSL}(2,\mathbb{C}), \; \phi \in (0,\pi).$$

The coefficients of matrices that commute or anti-commute with $\tilde{\rho}$ must fulfill $a = \pm\overline{d}$, $b = \pm\overline{c}e^{\mathrm{i}\phi} = \pm\overline{c}e^{-\mathrm{i}\phi}$. As $\phi \neq 0, \pi$, we have $b = c = 0$,

$$\begin{pmatrix} a & b \\ c & d \end{pmatrix} = \begin{pmatrix} a & 0 \\ 0 & \overline{a} \end{pmatrix}.$$

Let

$$\tilde{\xi} = \begin{pmatrix} e^{\mathrm{i}\pi/n} & 0 \\ 0 & e^{-\mathrm{i}\pi/n} \end{pmatrix}$$

be the smallest rotation of that kind, then the centralizer is the finite cyclic group $\mathcal{C}_\Theta(\tilde{\rho}) = \langle\tilde{\xi}\rangle = \mathcal{E}_\Theta(\tilde{\rho})$ and $\mathcal{C}_{\Gamma^+}(\rho) = \langle\xi\rangle = \mathcal{E}_{\Gamma^+}(\rho)$, $\xi = g^{-1}\tilde{\xi}g$.

We obtain

(63)
$$\frac{1}{2}\int\limits_{\mathcal{F}_{\mathcal{C}_{\Gamma^+}(\rho)}} d\mu(x)\; \chi(\rho)\; [G(\rho(x),x;E) - G(\rho(x),x;E')] = \frac{\chi(\rho)}{2\,\mathrm{ord}\mathcal{E}_{\Gamma^+}(\rho)}\,[I(p,\phi_\rho) - I(p',\phi_\rho)].$$

**Classes of inverse hyperbolic elements.** An inverse hyperbolic element $\tau$ is conjugate to

$$\tilde{\tau} = \begin{pmatrix} e^{l/2} & 0 \\ 0 & e^{-l/2} \end{pmatrix} \mathrm{j} = g\tau g^{-1}, \; g \in \mathrm{PSL}(2,\mathbb{C}), \; l > 0.$$

It follows from

$$\begin{pmatrix} e^{l/2} & 0 \\ 0 & e^{-l/2} \end{pmatrix} \mathrm{j} \begin{pmatrix} a & b \\ c & d \end{pmatrix} = \pm \begin{pmatrix} a & b \\ c & d \end{pmatrix} \begin{pmatrix} e^{l/2} & 0 \\ 0 & e^{-l/2} \end{pmatrix} \mathrm{j}$$

that

$$\begin{pmatrix} a & b \\ c & d \end{pmatrix} = \begin{pmatrix} e^{l'/2} & 0 \\ 0 & e^{-l'/2} \end{pmatrix} \quad \text{or} \quad \begin{pmatrix} a & b \\ c & d \end{pmatrix} = \begin{pmatrix} e^{l'/2} & 0 \\ 0 & e^{-l'/2} \end{pmatrix} \begin{pmatrix} \mathrm{i} & 0 \\ 0 & -\mathrm{i} \end{pmatrix}, \; l' > 0.$$

Let $\tilde{\tau}_0^+ \in \Theta = g\Gamma^+ g^{-1}$ be such an element with smallest occuring length $l_{\tau_0^+}$. The centralizer of $\tilde{\tau}$ is generated by that element and the maximal finite subgroup $\mathcal{E}_\Theta(\tilde{\tau})$ of $\mathcal{C}_\Theta(\tilde{\tau})$ :

(64)
$$\mathcal{C}_\Theta(\tilde{\tau}) = \langle\tilde{\tau}_0^+\rangle \times \mathcal{E}_\Theta(\tilde{\tau}),$$

(65)
$$\mathcal{E}_\Theta(\tilde{\tau}) = \left\{\begin{pmatrix} 1 & 0 \\ 0 & 1 \end{pmatrix}\right\} \quad \text{or} \quad \mathcal{E}_\Theta(\tilde{\tau}) = \left\{\begin{pmatrix} 1 & 0 \\ 0 & 1 \end{pmatrix}, \begin{pmatrix} \mathrm{i} & 0 \\ 0 & -\mathrm{i} \end{pmatrix}\right\}.$$

We integrate over the set

$$\left\{x_1, x_2 \in \mathbb{R}, 1 \le x_3 \le \exp\left(l_{\tau_0^+}\right)\right\},$$

which in the first case ($\text{ord}\mathcal{E}_\Theta(\tilde\tau) = \text{ord}\mathcal{E}_{\Gamma^+}(\tau) = 1$) is the fundamental cell of the centralizer, in the second case ($\text{ord}\mathcal{E}_\Theta(\tilde\tau) = \text{ord}\mathcal{E}_{\Gamma^+}(\tau) = 2$) just the double. We obtain

$$(66)\quad \frac{1}{2}\int\limits_{\mathcal{F}_{C_{\Gamma^+}(\tau)}} d\mu(x)\,\chi(\tau)\,[G(\tau(x),x;E) - G(\tau(x),x;E')]$$

$$= \frac{\chi(\tau)}{8\pi\,\text{ord}\mathcal{E}_{\Gamma^+}(\tau)} \int\limits_1^{\exp(l_{\tau_0^+})} \frac{dx_3}{x_3^3}\int\limits_{-\infty}^{\infty} dx_1 \int\limits_{-\infty}^{\infty} dx_2\, \frac{\exp[\mathrm{i}\,p\,d(\tau(x),x)] - \exp[\mathrm{i}\,p'\,d(\tau(x),x)]}{\sinh d(\tau(x),x)}\,,$$

where

$$\cosh d(\tau(x),x) = 1 + 2\,\frac{(x_1^2+x_3^2)\sinh^2 l/2 + x_2^2\cosh^2 l/2}{x_3^2}\,.$$

After having performed the integration with the help of [25] we get a contribution

$$(67)\qquad -\frac{\chi(\tau)}{8\,\text{ord}\mathcal{E}_{\Gamma^+}(\tau)\sinh l}\left(\frac{\exp(\mathrm{i}\,p\,l_\tau)}{\mathrm{i}p} - \frac{\exp(\mathrm{i}\,p'\,l_\tau)}{\mathrm{i}p'}\right).$$

Collecting all contributions we get a formula for the trace of the regularized resolvent for $\text{Im}\,p, \text{Im}\,p' > 1$

$$\sum_{n=1}^{\infty}\left[(p_n^2 - p^2)^{-1} - (p_n^2 - p'^2)^{-1}\right]$$

$$(68)\qquad = -\frac{\text{Vol}(\mathcal{F}_{\Gamma^+})}{8\pi\mathrm{i}}(p - p')$$

$$-\sum_{\substack{\{\rho\}_{\Gamma^+}\\ \text{inv.}}} \chi(\sigma)\,\frac{\text{Area}(\mathcal{P}_\rho)}{16\pi}\,[\psi(1-\mathrm{i}p) + \psi(-\mathrm{i}p) - \psi(1-\mathrm{i}p') - \psi(-\mathrm{i}p')]$$

$$-\sum_{\substack{\{\rho\}_{\Gamma^+}\\ \text{ellipt.}}} \frac{\chi(\rho)\,l_{\tau_0^+}}{8\,\text{ord}\mathcal{E}_{\Gamma^+}(\rho)\,(1-\cos\phi_\rho)}\left(\frac{1}{\mathrm{i}p} - \frac{1}{\mathrm{i}p'}\right)$$

$$+\sum_{\substack{\{\rho\}_{\Gamma^+}\\ \text{inv. ellipt.}}} \frac{\chi(\rho)}{2\,\text{ord}\mathcal{E}_{\Gamma^+}(\rho)}[I(p,\phi_\rho) - I(p',\phi_\rho)]$$

$$-\sum_{\substack{\{\tau\}_{\Gamma^+}\\ \text{hyperbol.}}} \frac{\chi(\tau)\,l_{\tau_0^+}}{8\,\text{ord}\mathcal{E}_{\Gamma^+}(\tau)\,(\cosh l_\tau - \cos\phi_\tau)}\left(\frac{\exp(\mathrm{i}\,p\,l_\tau)}{\mathrm{i}p} - \frac{\exp(\mathrm{i}\,p'\,l_\tau)}{\mathrm{i}p'}\right)$$

$$-\sum_{\substack{\{\tau\}_{\Gamma^+}\\ \text{inv. hyperbol.}}} \frac{\chi(\tau)\,l_{\tau_0^+}}{8\,\text{ord}\mathcal{E}_{\Gamma^+}(\tau)\,\sinh l_\tau}\left(\frac{\exp(\mathrm{i}\,p\,l_\tau)}{\mathrm{i}p} - \frac{\exp(\mathrm{i}\,p'\,l_\tau)}{\mathrm{i}p'}\right).$$

To put our trace formula into the form (14), we will rewrite the sums over $\Gamma^+$-conjugacy classes as sums over $\Gamma$-conjugacy classes:

**Case 1.** Suppose $g \in \Gamma$ is a plane reflection. Then

$$\{g\}_\Gamma = \{g\}_{\Gamma^+} \cup \{g\}_{\Gamma^+\sigma} = \{g\}_{\Gamma^+},$$

as $\{g\}_{\Gamma^+} = \{g\}_{\Gamma^+\sigma} = \{\sigma g\sigma\}_{\Gamma^+}$, i.e., there exists an $h \in \Gamma^+$, such that $g = h\sigma g\sigma h^{-1}$. For example, choose $h = g\sigma$.

**Case 2.** Suppose $g \in \Gamma$ is not a plane reflection and the decomposition

$$\{g\}_\Gamma = \{g\}_{\Gamma^+} \cup \{g\}_{\Gamma^+\sigma} = \{g\}_{\Gamma^+} \cup \{\sigma g \sigma\}_{\Gamma^+}$$

is disjoint. Then there does not exist an $h \in \Gamma^+$, such that $g = h\sigma g \sigma h^{-1} = h\sigma g (h\sigma)^{-1}$. In this case no element of $\Gamma^+\sigma$ commutes with $g$, therefore $\mathcal{C}_{\Gamma^+}(g) = \mathcal{C}_\Gamma(g)$. Furthermore $\mathcal{E}_\Gamma(g) = \mathcal{E}_{\Gamma^+}(g)$, in particular $\mathrm{ord}\mathcal{E}_\Gamma(g) = \mathrm{ord}\mathcal{E}_{\Gamma^+}(g) = \mathrm{ord}\mathcal{E}_{\Gamma^+}(\sigma g \sigma)$. The last equality holds, as $\mathcal{E}_{\Gamma^+}(g) = \sigma\, \mathcal{E}_{\Gamma^+}(\sigma g \sigma)\, \sigma$ and conjugate groups have the same order.

Let $\tau_0^+$ be a hyperbolic element with shortest length commuting with $g$. *In this case $\tau_0^+$ is not only $\Gamma^+$-primitive, but also $\Gamma$-primitive.*

We show this by contradiction. Suppose $\tau_0^+$ is not $\Gamma$-primitive. Then there exists a $\tau_0 \in \Gamma^+\sigma$, such that $\tau_0^2 = \tau_0^+$. Conjugation by $\tau_0$ yields $\tau_0^2 = \tau_0 \tau_0^+ \tau_0^{-1}$, hence $\tau_0$ commutes with $\tau_0^+$ – contradiction!

As $\tau_0^+$ is $\Gamma$-primitive, we write $\tau_0 \equiv \tau_0^+$. We abbreviate the summands in (68) by

$$\frac{l_{\tau_0^+} S(l_g, \phi_g)}{\mathrm{ord}\mathcal{E}_{\Gamma^+}}$$

and obtain

$$\frac{2 l_{\tau_0} S(l_g, \phi_g)}{\mathrm{ord}\mathcal{E}_\Gamma(g)} = \frac{l_{\tau_0^+} S(l_g, \phi_g)}{\mathrm{ord}\mathcal{E}_{\Gamma^+}(g)} + \frac{l_{\tau_0^+} S(l_g, \phi_g)}{\mathrm{ord}\mathcal{E}_{\Gamma^+}(\sigma g \sigma)}. \tag{69}$$

Now $S(l_g, \phi_g) = S(l_g, -\phi_g)$ and, as $-\mathrm{j}\,(\mathrm{tr} g)\,\mathrm{j} = \overline{\mathrm{tr} g}$, we get

$$S(l_g, -\phi_g) = S(l_{\mathrm{j}g\mathrm{j}}, \phi_{\mathrm{j}g\mathrm{j}}) = S(l_{\sigma g \sigma}, \phi_{\sigma g \sigma}).$$

It follows

$$\frac{2 l_{\tau_0} S(l_g, \phi_g)}{\mathrm{ord}\mathcal{E}_\Gamma(g)} = \frac{l_{\tau_0^+} S(l_g, \phi_g)}{\mathrm{ord}\mathcal{E}_{\Gamma^+}(g)} + \frac{l_{\tau_0^+} S(l_{\sigma g \sigma}, \phi_{\sigma g \sigma})}{\mathrm{ord}\mathcal{E}_{\Gamma^+}(\sigma g \sigma)}. \tag{70}$$

**Case 3.** Suppose $g \in \Gamma$ is not a plane reflection and the decomposition

$$\{g\}_\Gamma = \{g\}_{\Gamma^+} \cup \{g\}_{\Gamma^+\sigma}$$

is not disjoint. Then $\{g\}_{\Gamma^+} = \{g\}_{\Gamma^+\sigma}$, i.e., there must be an $h \in \Gamma^+$, such that $g = h\sigma g (h\sigma)^{-1}$. Hence the centralizer $\mathcal{C}_\Gamma(g)$ contains orientation reversing elements. $\mathcal{C}_{\Gamma^+}(g)$ is the subgroup of all orientation preserving elements. It is of index two in $\mathcal{C}_\Gamma(g)$:

$$\mathcal{C}_\Gamma(g) = \mathcal{C}_{\Gamma^+}(g) \,\dot\cup\, \mathcal{C}_{\Gamma^+}(g)\hat\sigma.$$

If $\mathcal{C}_{\Gamma^+}(g)\hat\sigma$ *contains elements of finite order*, $\hat\sigma$ can be chosen such that it is itself of finite order,

$$\mathcal{E}_\Gamma(g) = \mathcal{E}_{\Gamma^+}(g) \,\dot\cup\, \mathcal{E}_{\Gamma^+}(g)\hat\sigma.$$

*In this case, a hyperbolic element $\tau_0^+$ with shortest length that commutes with $g$ is $\Gamma$-primitive.*

Suppose the contrary, i.e., there is a $\tau_0 \in \Gamma - \Gamma^+$ with $\tau_0^2 = \tau_0^+$. But now $\tau_0\hat\sigma$ is hyperbolic having the same length as $\tau_0$, hence half the length of $\tau_0^+$. The latter should have been a hyperbolic element with shortest length – contradiction.

We obtain

$$
\frac{2l_{\tau_0} S(l_g, \phi_g)}{\mathrm{ord}\mathcal{E}_\Gamma(g)} = \frac{l_{\tau_0^+} S(l_g, \phi_g)}{\mathrm{ord}\mathcal{E}_{\Gamma^+}(g)}. \tag{71}
$$

If $\mathcal{C}_{\Gamma^+}(g)\hat{\sigma}$ *contains no elements of finite order*, then

$$
\mathcal{E}_\Gamma(g) = \mathcal{E}_{\Gamma^+}(g).
$$

We show *there exists a hyperbolic element* $\tau_1$ *with shortest length in* $\mathcal{C}_{\Gamma^+}(g)$*, which is not* $\Gamma$-*primitive* (so there is an element $\tau_0$ in $\mathcal{C}_\Gamma(g)$ with a shorter length):

Let $\tau_0^+$ be a hyperbolic element with shortest length $l_{\tau_0^+}$ in $\mathcal{C}_{\Gamma^+}(g)$ and $\tau_0$ an inverse hyperbolic element with shortest length $l_{\tau_0}$ in $\mathcal{C}_{\Gamma^+}(g)\hat{\sigma}$.

$l_{\tau_0^+}$ cannot be smaller than $l_{\tau_0}$, because otherwise either $\tau_0\tau_0^+ \in \mathcal{C}_{\Gamma^+}(g)\hat{\sigma}$ or $\tau_0{\tau_0^+}^{-1} \in \mathcal{C}_{\Gamma^+}(g)\hat{\sigma}$ would have a smaller length than $\tau_0$, which already should have had the smallest length. $l_{\tau_0^+} = l_{\tau_0}$ is not possible either, because otherwise $\tau_0\tau_0^+ \in \mathcal{C}_{\Gamma^+}(g)\hat{\sigma}$ or $\tau_0{\tau_0^+}^{-1} \in \mathcal{C}_{\Gamma^+}(g)\hat{\sigma}$ would be of finite order. Hence $l_{\tau_0^+} = 2l_{\tau_0}$. $\tau_1 := \tau_0^2$ is obviously not $\Gamma$-primitive. Therefore

$$
\frac{2l_{\tau_0} S(l_g, \phi_g)}{\mathrm{ord}\mathcal{E}_\Gamma(g)} = \frac{l_{\tau_0^+} S(l_g, \phi_g)}{\mathrm{ord}\mathcal{E}_{\Gamma^+}(g)}. \tag{72}
$$

The above arguments allow to rewrite formula (68) in the form (14).

# Acknowledgment

We would like to thank Professor F. Steiner for many helpful discussions and the Deutsche Forschungsgemeinschaft for financial support.